\documentclass[10pt, conference, compsocconf]{IEEEtran}
\pdfoutput=1

\newcommand{\ignore}[1]{}
\usepackage{amsmath}
\usepackage{amssymb}
\usepackage{graphicx}
\usepackage{cite}
\usepackage{comment}
\usepackage{algorithm}
\usepackage{algorithmicx}
\usepackage{algpseudocode}
\usepackage{array}
\usepackage{url}
\usepackage{epstopdf}

\DeclareGraphicsExtensions{.eps,.pdf}
\graphicspath{{.}}

\begin{document}
%
\title{Testing fine-grained parallelism for the ADMM on a factor-graph}

\author{\IEEEauthorblockN{Ning Hao}
\IEEEauthorblockA{ning.hao@oracle.com}
\and
\IEEEauthorblockN{AmirReza Oghbaee}
\IEEEauthorblockA{oghbaeea@ece.neu.edu}
\and
\IEEEauthorblockN{Mohammad Rostami}
\IEEEauthorblockA{mrostami@seas.upenn.edu}
\and
\IEEEauthorblockN{Nate Derbinsky}
\IEEEauthorblockA{derbinskyn@wit.edu}
\and
\IEEEauthorblockN{Jos\'e Bento}
\IEEEauthorblockA{jose.bento@bc.edu}
}

\maketitle

\begin{abstract}
There is an ongoing effort to develop tools that apply distributed computational resources to tackle large problems or reduce the time to solve them.
In this context, the Alternating Direction Method of Multipliers (ADMM) arises as a method that can exploit distributed resources like the dual ascent method
and has the robustness and improved convergence of the augmented Lagrangian method.
%
Traditional approaches to accelerate the ADMM using multiple cores are problem-specific and often require multi-core programming.
%
By contrast, we propose a problem-independent scheme of accelerating the ADMM that does not require the user to write any parallel code.
We show that this scheme, an interpretation of the ADMM as a message-passing algorithm on a factor-graph, can automatically exploit fine-grained parallelism both in GPUs and shared-memory multi-core
computers and achieves significant speedup in such diverse application domains as combinatorial optimization, machine learning, and optimal control.
%
%
%
Specifically, we obtain 10-18x speedup using a GPU, and 5-9x using multiple CPU cores, over a serial, optimized C-version of the ADMM, \ignore{also available with our tool,} which is similar to the typical speedup reported for existing GPU-accelerated libraries, including cuFFT (19x), cuBLAS (17x), and cuRAND (8x).
\end{abstract}


\IEEEpeerreviewmaketitle

\section{Introduction}

\noindent
The Alternating Direction Method of Multipliers (ADMM) is 
a popular iterative algorithm to solve non-smooth optimization problems in a distributed way. 
The algorithm is often presented as solving problems of the form
\begin{align} \label{eq:classical_admm_form}
&\text{minimize } f(w_1) + g(w_2)\nonumber\\
&\text{subject to } Aw_1 + B w_2 = c
\end{align}
using the following iterations on auxiliary variables $x,z,u$:
\begin{algorithm}[h!]
\caption{ADMM}
\begin{algorithmic}[1] 
\While{!\text{stopping criteria}}
\State $x \leftarrow \arg \min_s  f(s) + \frac{\rho}{2} \|A s + B z - c + u \|^2 $\label{eq:x_update_classical_ADMM}
\State $z \leftarrow \arg \min_r  g(r) + \frac{\rho}{2} \|A x + B r - c + u \|^2 $\label{eq:z_update_classical_ADMM}
\State $u \leftarrow u  + Ax + Bz  - c$ \label{eq:u_update_classical_ADMM}
\EndWhile
\end{algorithmic} \label{alg:ADMM_classical_presentation}
\end{algorithm}

\noindent
After convergence, the optimal solution $(w^*_1,w^*_2)$ can be read from $(x,z)$.
Convergence is guaranteed under convexity and some mild technical assumptions\cite{Boyd2011} although it has been used with surprising success for non-convex applications ranging from computer graphics to power systems\cite{erseghe2014distributed, mota2012distributed, zoran2014shape}.
The free parameter $\rho > 0$ can be used to control convergence. 

One opportunity to exploit distributed computational resources with the ADMM arises when the functions $f$ and/or $g$ and the matrices $A$ and/or $B$ make the subproblems in lines \ref{eq:x_update_classical_ADMM}--\ref{eq:z_update_classical_ADMM} of Algorithm \ref{alg:ADMM_classical_presentation} decomposable into smaller independent problems. 
As a simple example, consider $f(w_1) = f_1(w_{11}) + f_2(w_{12}) + f_3(w_{13})$ and let $A$ be the identity matrix.  
Line \ref{eq:x_update_classical_ADMM} then decomposes into three independent updates for $w_{11}$, $w_{12}$, and $w_{13}$ involving the functions $f_1$, $f_2$, and $f_3$, respectively.
The authors in \cite{Boyd2011} use this idea to decompose a Lasso problem involving $30$GB of data and $8000$ features into $80$ sub-problems, each solved by a separate computer.

Another opportunity to exploit distributed resources arises when lines \ref{eq:x_update_classical_ADMM}--\ref{eq:u_update_classical_ADMM} involve operations that can be accelerated using a GPU and/or multiple CPU-cores.
A good example of this is \cite{bhaskar2011admm}, where a GPU and the JACKET toolbox for MATLAB are applied to accelerate matrix multiplications in an ADMM-based solution to the sparse coding problem. 
Unlike the first scenario, this approach is not specific to the ADMM (e.g. GPUs are commonly used to speed linear-algebra computations when fitting neural nets to data using stochastic gradient descent).

In both scenarios, exploiting parallelism only becomes evident \textit{after} a problem is specified, which explains why most GPU-accelerated implementations of the ADMM have been problem specific.
Here we numerically study a \textit{problem-independent} approach and we observe, across multiple tasks, that this generality does not preclude useful acceleration.

This result is most surprising in the context of typical GP-GPU acceleration (e.g. linear algebra), in which GPU cores are assigned relatively simple and similar tasks.
By contrast, our parallelization scheme schedules relatively complex and dissimilar tasks, and yet still performs quite well.

We also study the performance of our framework using multiple CPU cores, thus providing new data to answer the question raised by \cite{lee2010debunking} of whether GPUs are really game-changing or whether we are better off simply exploiting the multi-core parallelism of modern CPUs.

Due to these positive findings, we have developed \textit{parADMM}, an open-source, general-purpose optimization tool based on the ADMM that \textbf{allows users to automatically exploit GPU and multi-CPU parallelism}. 
To the best of our knowledge, there is no other GPU-accelerated, general-purpose ADMM solver as versatile and automatic to use.
At the time of writing, the only comparable tool is SNAPVX\cite{hallac2015snapvx}, which (a) is written in Python and hence much slower; (b) forces users to use CVXPY\cite{diamond2015cvxpy}, adding additional delay; (c) is restricted to convex problems; (d) can only solve problems that decompose into a very particular form; and (e) can only exploit multiple CPU-cores and not GPU parallelism. {\it parADMM} does not share these limitations.

\section{The ADMM on a factor-graph}

\noindent
Our starting point is a formulation of the ADMM for a factor-graph representation of the objective function that makes explicit many small operations that can be performed in parallel; hence \textit{fine-grained} parallelism.

First we write the objective function as
\begin{equation} \label{eq:gen_opt_prob}
f(w) = \sum_{a\in F} f_a(w_{\partial a}),
\end{equation}
where $w = (w_1,\ldots,w_p)\in \mathbb{R}^{p \times d}$ are variables, $\partial a \subseteq V \equiv [p] \equiv \{1,\ldots,p\}$ is a subset of indices, and
$w_{\partial a} = \{w_i:i \in \partial a\}$ is a subset of variables.
Functions $\{f_a\}_{a\in F}$ take values in $\mathbb{R} \cup \{\pm\infty\}$ and do not need to be smooth.
Thus \eqref{eq:gen_opt_prob} includes constrained optimization, and
\eqref{eq:classical_admm_form} and \eqref{eq:gen_opt_prob} are equally general.
%

\begin{figure}[b!]
\begin{center}
\includegraphics[scale=0.3]{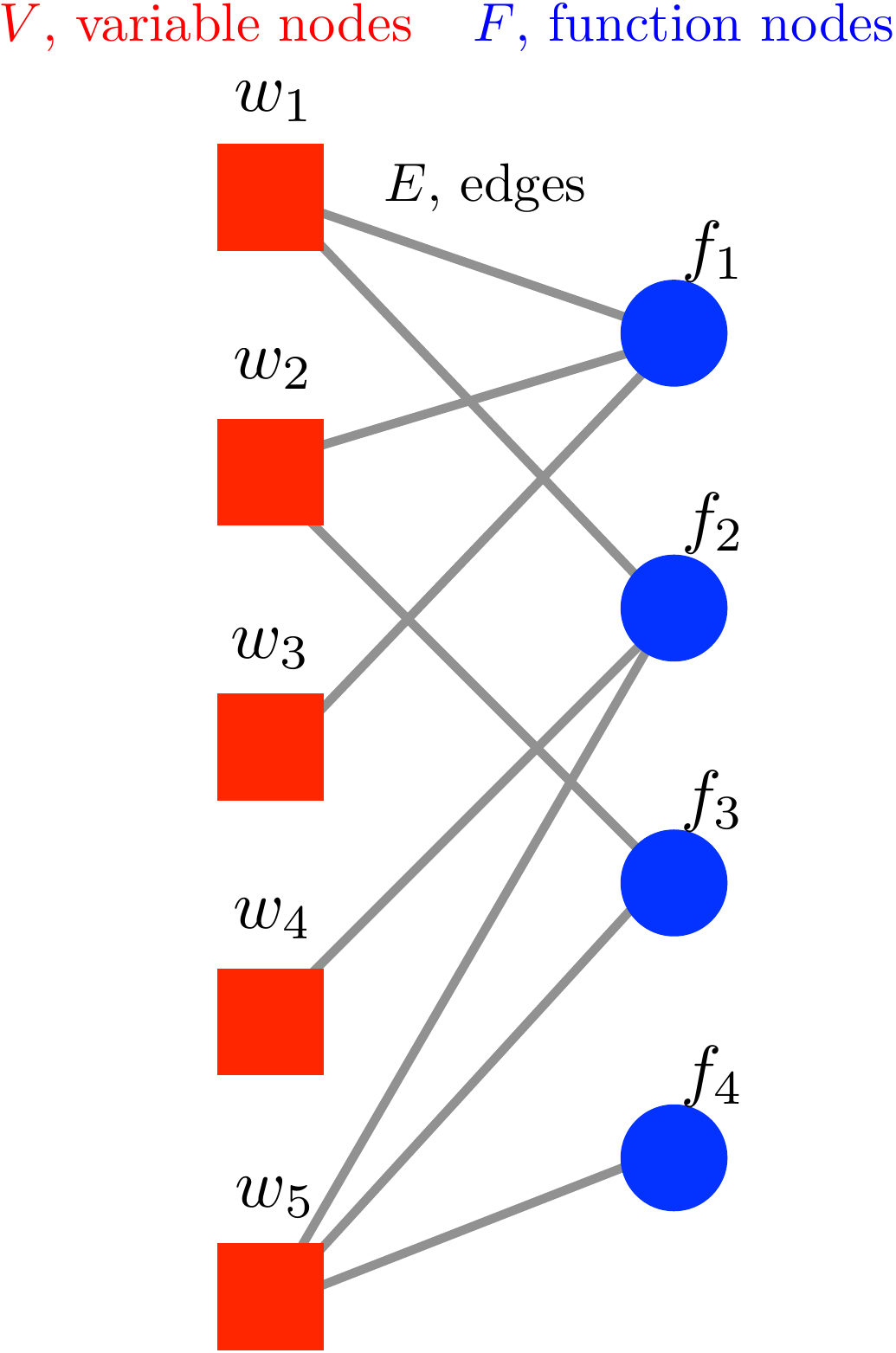} \;\;\;\;\;\;\;\;\;
\includegraphics[scale=0.3]{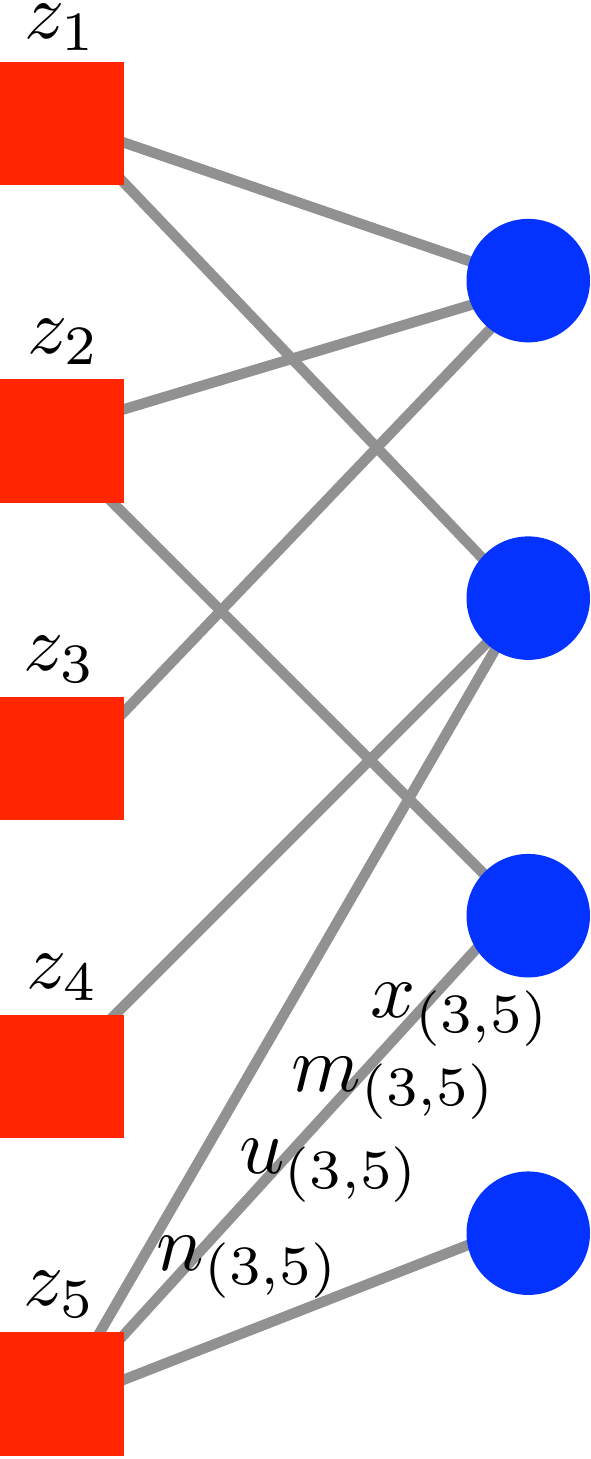}
\caption{Left - Factor-graph representation of an objective function; \hspace{1.0cm} Right - Auxiliary variables in the message-passing ADMM.}
\vspace{-4mm}
\label{fig:factor_graph}
\end{center}
\end{figure}

This objective function can be written as a factor-graph $G = (F,V,E)$ where
edge $(a,b) \in E$ represents a dependency of function $f_a \in F$ on the component $w_b \in V$.
Figure \ref{fig:factor_graph}-left shows this
factor-graph representation of $f(w) = f_1(w_1,w_2,w_3) + f_2(w_1,w_4,w_5) + f_3(w_2,w_5) + f_4(w_5)$.

The message-passing ADMM is a scheme that updates five auxiliary variables $x,m,z,u,n$ on the graph $G$.
We can interpret these updates as a message-passing scheme\cite{Derbinsky2013AnImp}.
For each $a\in F$ we define the neighbors of $a$ as $\partial a = \{b\in V: (a,b)\in E\}$; these are the components of $w$ that function $f_a$ depends on.
For each $b \in V$ we define the neighbors of $b$ as $\partial b = \{a\in F: (a,b)\in E\}$; these are the functions that depend on component $w_b$.
We denote the number of elements in $\partial a$ and $\partial b$ as $|\partial a|$ and $|\partial b|$, respectively.

The relationship between the ADMM auxiliary variables $x,m,u,n,z$ and the factor-graph is exemplified in Figure \ref{fig:factor_graph}-right and is as follows.
Each edge $(a,b) \in E$ is associated with four variables $x_{(a,b)},m_{(a,b)},u_{(a,b)}$, and $n_{(a,b)}$.
Each variable node $b \in V$ is associated with one variable $z_b$ and each function node $a \in F$ is associated with one function $f_a$.
Each edge $(a,b)\in E$ is also associated with two parameters $\rho_{(a,b)},\alpha_{(a,b)} > 0$, which in classical implementations are considered constant but for which there are also improved update schemes (e.g. \cite{Derbinsky2013AnImp} which {\it parADMM} can also implement).
The free parameters $\rho$ and $\alpha$ allow us to control the convergence rate of the algorithm.

\begin{algorithm}[t!]
\caption{ADMM on a factor-graph}
\begin{algorithmic}[1]
\While{!\text{stopping criteria}}
\For{$a \in F$} \Comment{$x$-update}
\State $x_{(a,\partial a)} \hspace{-1mm}\leftarrow \text{\bf Prox}_{f_a,\rho_{(a,b)}}(n_{(a,b)})$ \label{eq:x_update}
\EndFor
\For{$(a,b) \in E$}  \Comment{$m$-update}
\State $m_{(a,b)} \leftarrow x_{(a,b)} + u_{(a,b)}$
\EndFor
\For{$b\in V$}  \Comment{$z$-update}
\State $z_{b} \leftarrow \sum_ {a \in \partial b} \rho_{(a,b)} m_{(a,b)} \Big / \sum_ {a \in \partial b} \rho_{(a,b)}$ \label{eq:z_update}
\EndFor
\For{$(a,b) \in E$}  \Comment{$u$-update}
\State $u_{(a,b)} \leftarrow u_{(a,b)} + \alpha_{(a,b)} (x_{(a,b)} - z_b )$
\EndFor
\For{$(a,b) \in E$}  \Comment{$n$-update}
\State $n_{(a,b)} \leftarrow z_{b} - u_{(a,b)}$ \label{eq:n_update}
\EndFor
\EndWhile
\end{algorithmic}  \label{alg:ADMM}
\end{algorithm}

The ADMM updates these variables sequentially and in a cyclic way as described in Algorithm \ref{alg:ADMM}.
To shorten the notation, given a function $h$ and a constant $\rho>0$, we denote by $\text{\bf Prox}_{h,\rho}(r)$ the map from $r$ to $s$ defined by
\begin{equation} \label{eq:PO}
\arg \min_{s} h(s) + \frac{\rho}{2} \|s -  r\|^2.
\end{equation}
This is termed the proximal operator (PO) of $h$.
After a fixed number of iterations, or a desired accuracy is achieved, the solution $w^*$ is read from the variables $z$.

The advantage of Algorithm \ref{alg:ADMM} over Algorithm \ref{alg:ADMM_classical_presentation} is that Algorithm \ref{alg:ADMM} utilizes five for-loops that can be parallelized independent of the target problem.
We exploit this property by assigning calculation of each PO to a different GPU/CPU core, allowing users to exploit parallelism while only writing serial code to compute each PO.

\section{parADMM}

\noindent
We use {\it parADMM} to study our parallelization scheme on GPUs and CPUs, and so it is useful to detail how it works.
We also invite the reader to use {\it parADMM} to solve other problems
(\url{https://github.com/parADMM/engine}).
Later we evaluate this parallel scheme in three application domains.

Most of {\it parADMM} is written in C for speed.
A portion is written in CUDA to exploit GPU parallelism and another portion makes uses of OpenMP to exploit multi CPU-core parallelism in shared-memory systems.
The limits of the current version are the computer memory and the GPU memory; however, many real-life problems can easily fit in a single large-memory server or on a GPU. Right now we can solve problems with factor-graphs involving millions of nodes. The extension to multi-computer and multi-GPU is not fundamentally challenging, but is still in the pipeline.

Our tool supports any kind of factor-graph and PO.
However, we observe the greatest acceleration when
\begin{enumerate}
\item the factor-graph is large (i.e. the user decomposes the problem into many sub-problems); and/or
\item if using a GPU, the sub-problems are relatively simple (i.e. the serial code for each PO is not too complex). Nonetheless, all POs we evaluate contain code that is substantially more complex than is typical in GPU-accelerated libraries, such as linear algebra routines.
\end{enumerate}
%

\begin{figure}[t!]
\begin{center}
\includegraphics[trim=0.8cm 2cm 0cm 0cm, clip=true,scale=0.65]{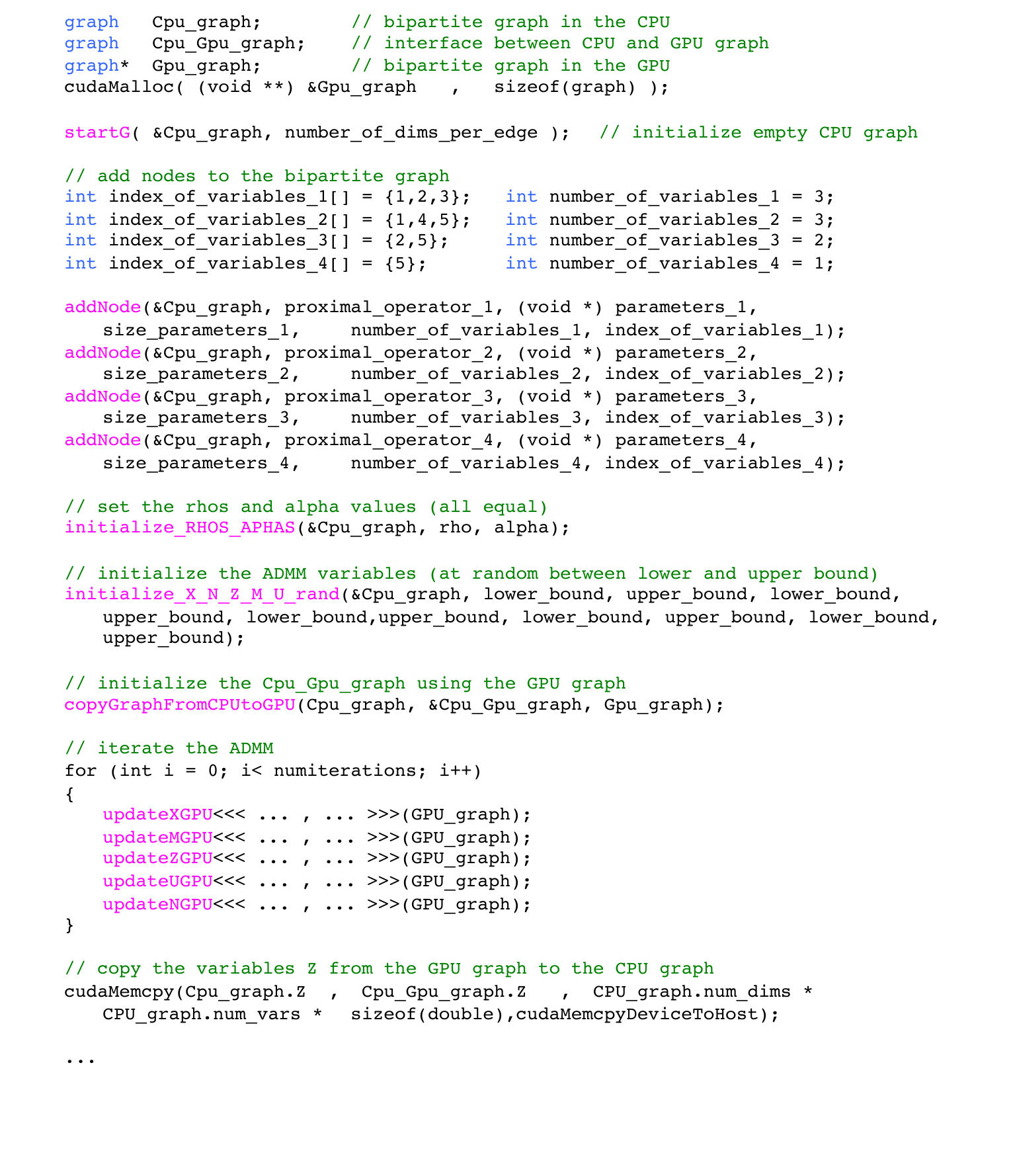}
\caption{General program structure to solve the problem in Figure \ref{fig:factor_graph}.\vspace{-0.4cm}}
\vspace{-3mm}
\label{fig:general_code}
\end{center}
\end{figure}

We focus on solving problems with GPU first.
In Figure \ref{fig:general_code} we illustrate how to use {\it parADMM} to solve the problem in Figure \ref{fig:factor_graph}.
The two main tasks to solve a problem are
\begin{enumerate}
\item specifying the topology of the factor-graph via the {\fontfamily{pcr}\selectfont addNode} function; and
\item providing {\bf serial code} to compute each PO, i.e. solve problem \eqref{eq:PO}, via function pointers, e.g. {\fontfamily{pcr}\selectfont proximal\_operator\_1}.
\end{enumerate}

\noindent
The variable {\fontfamily{pcr}\selectfont Cpu\_graph} is a structured C variable stored in the CPU that encodes $G = (V,F,E)$, as well as all ADMM auxiliary variables: $x,m,z,u,n$.
The function {\fontfamily{pcr}\selectfont addNode} extends {\fontfamily{pcr}\selectfont Cpu\_graph} by one node.
For example, the first call to {\fontfamily{pcr}\selectfont addNode} connects the PO {\fontfamily{pcr}\selectfont proximal\_operator\_1} to variable nodes $1,2,3$, specified via {\fontfamily{pcr}\selectfont index\_of\_variables\_1},
and also supplies any necessary parameters via {\fontfamily{pcr}\selectfont parameters\_1}.

To run the ADMM on the GPU we need to copy {\fontfamily{pcr}\selectfont Cpu\_graph} from the CPU to the GPU's global memory via the {\fontfamily{pcr}\selectfont CopyGraphFromCPUtoGPU} procedure, which automatically copies all parameters associated with the PO to the GPU global memory.
With this functionality, the user {\bf does not have to deal directly with GPU memory}, unless s/he so desires.
The factor-graph in GPU global memory is then referenced by the GPU pointer {\fontfamily{pcr}\selectfont Gpu\_graph}.

The main loop consists of calling five CUDA kernels to update each of the five different kinds of variables in parallel.
The two parameters inside each {\fontfamily{pcr}\selectfont <<<...,...>>>} are the number of blocks, $nb$, and the number of threads per block, $ntb$.
Since each graph variable is updated on a different core, the quantity $nb \times ntb$ must be just larger than $|F|$, $|E|$, $|V|$, $|E|$, and $|E|$ for the $x$-update, $m$-update, $z$-update, $u$-update, and $n$-update kernel calls, respectively.
Thus, once $ntb$ is specified, $nb$ is easily fixed.
As with most GPU applications, the performance of the tool depends upon the relative values of $nb$ and $ntb$ (c.f. Numerical Results).
Although GPU vendors like NVIDIA suggest that we make $ntb$ as large as possible (i.e. $ntb = 1024$), we find that for our scheme using a smaller number of threads-per-block gives better results.
Most of the time, we use $ntb = 32$, the smallest possible sensible value.
The work of \cite{volkov2010better} has good examples and justifications of when/why to choose a small $ntb$.

\begin{figure}[b!]
\begin{center}
\includegraphics[trim=0.2cm 0cm 0cm 0cm, clip=true,scale=0.65]{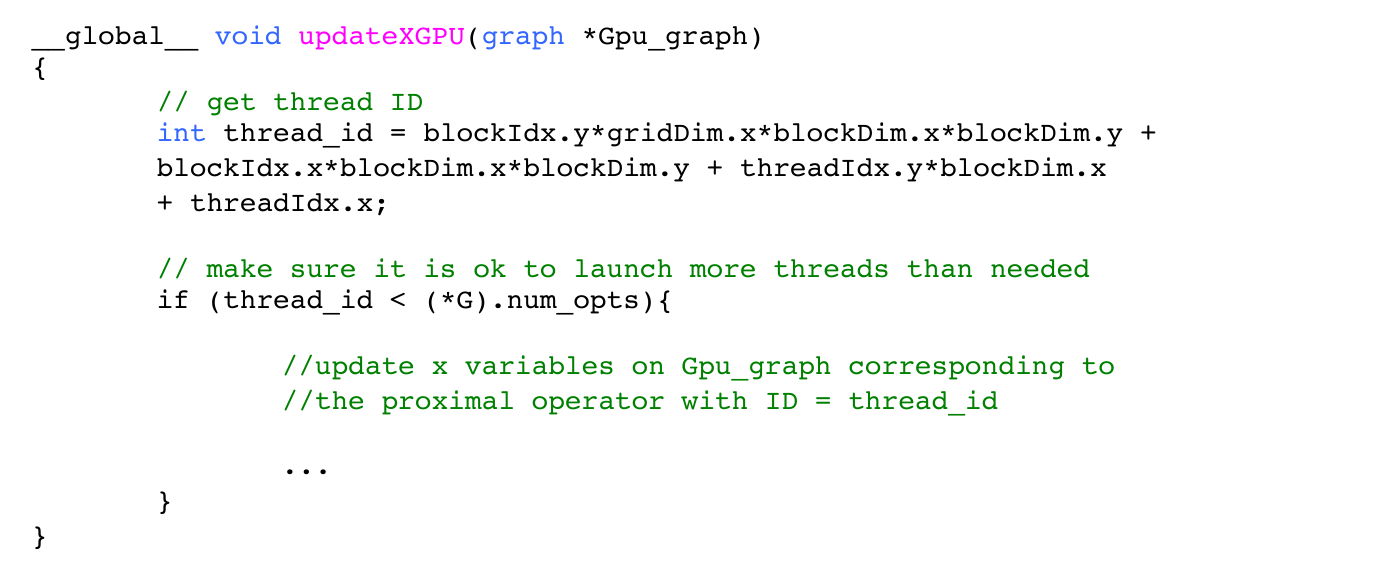}
\caption{CUDA kernel to update the $x$ variables.}
\vspace{-3mm}
\label{fig:GPU_kernel_update_x}
\end{center}
\end{figure}

The five GPU kernels share a similar structure.
Each GPU thread updates the variables in the factor-graph associated with just {\bf one} graph element: the $x$ variables of {\bf one} PO, the $m$ variables of {\bf one} edge, etc.
Figure \ref{fig:GPU_kernel_update_x} shows the structure of the kernel of the {\fontfamily{pcr}\selectfont updateXGPU<<<...,...>>>}.


All ADMM auxiliary variables are stored in 1-D arrays of doubles in GPU global memory (e.g. {\fontfamily{pcr}\selectfont Gpu\_graph.x} stores all the $x$ variables).
Variables $x,m,u,n$ are stored in their respective arrays in the order in which the respective edges where created.
For the code in Figure \ref{fig:general_code}, {\fontfamily{pcr}\selectfont Gpu\_graph.x} $= [x_{(1,1)}, x_{(1,2)}, x_{(1,3)}, x_{(2,1)}, x_{(2,4)}, x_{(2,5)}, x_{(3,2)}, x_{(3,5)}, x_{(4,5)}]$, which has $9 \times${\fontfamily{pcr}\selectfont number\_of\_dims\_per\_edge} doubles.
However, the $z$ variables are stored in {\fontfamily{pcr}\selectfont Gpu\_graph.z} in the order in which the variables were introduced into the problem.
For the code in Figure \ref{fig:general_code}, {\fontfamily{pcr}\selectfont Gpu\_graph.z} $= [z_{1}, z_{2}, z_{3}, z_{4}]$, which has $4 \times${\fontfamily{pcr}\selectfont number\_of\_dims\_per\_edge} doubles.

When using GPUs the user can control the order in which the variables are stored in the different arrays to exploit memory coalescence and achieve greater acceleration. In the current version of {\it parADMM} this order is a function of the sequence of node-additions performed.
For example, in an ideal scenario all threads in a thread-block are applying the same PO map to blocks of variables in sequence.
In a less ideal scenario, threads apply totally different POs to non-consecutive memory positions.

\subsection{Shared-memory multi-processor machines}

\noindent
\cite{lee2010debunking} reports that in many tasks, acceleration from a GPU is not much better than using modern multi-core CPUs and criticizes papers that advocate for the use of GPUs, but do not compare GPU vs. multiple CPU-cores.
In this paper we compare multiple CPU vs. GPU to see the degree to which this claim applies to our work; for this purpose we make it easy to exploit multiple CPUs in a shared-memory system using {\it parADMM}.
To do so we use OpenMP \cite{dagum1998openmp} to automatically perform each graph element update completely inside a core of a multi-CPU system: as with GPUs, the user's program is accelerated without writing any parallel code.
The user just removes {\fontfamily{pcr}\selectfont copyGraphFromCPUtoGPU()} and changes GPU updates to OpenMP (e.g. {\fontfamily{pcr}\selectfont updateXGPU<<<\ldots>>>(\ldots)} to {\fontfamily{pcr}\selectfont updateXOpenMP(\ldots)}).

\begin{figure}
\begin{center}
\includegraphics[trim=0cm 0cm 0cm 0cm, clip=true,scale=0.5]{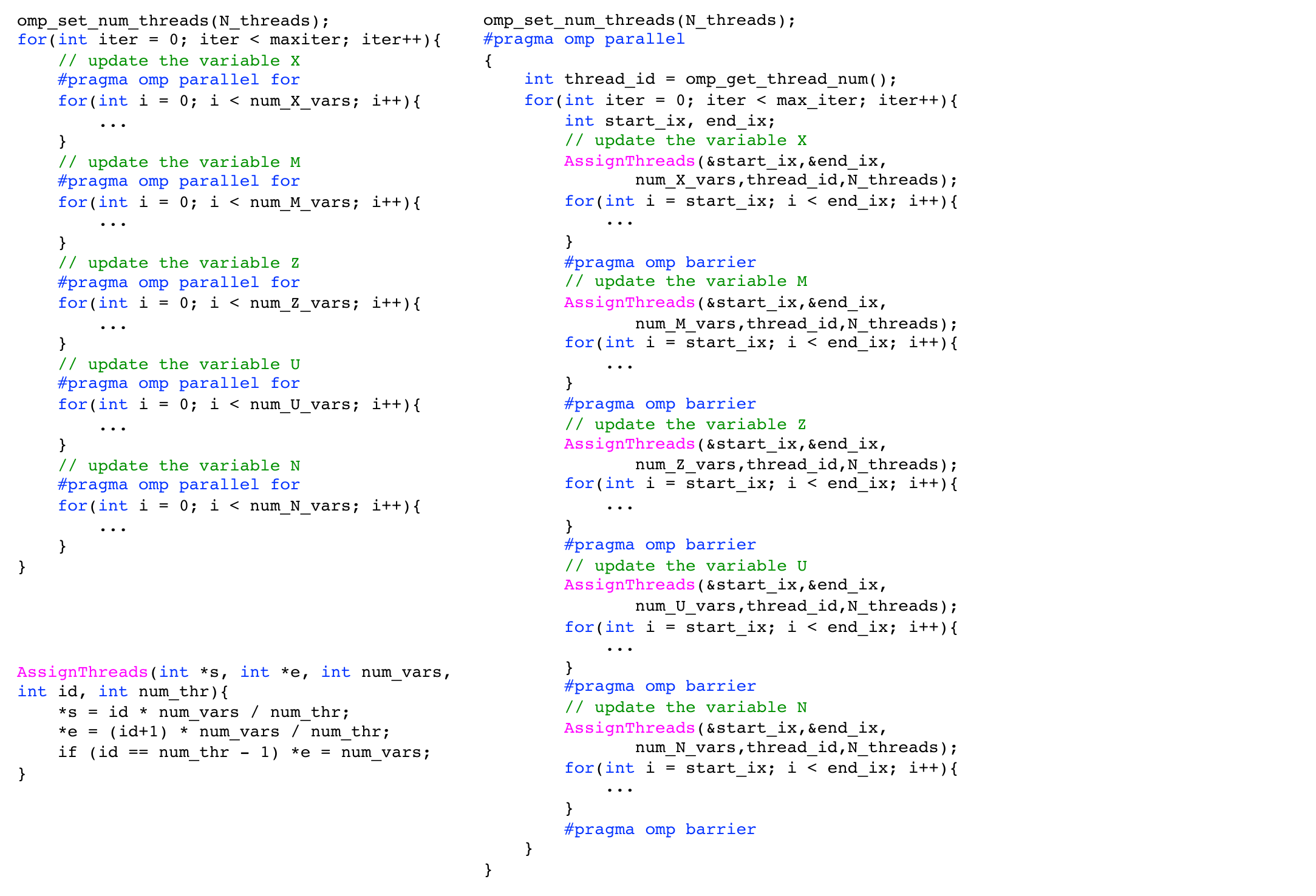}
\caption{First OpenMP approach (top-left) vs. second (bottom-left \& right). The first approach was faster in all the three problems tested.}
\vspace{-3mm}
\label{fig:openmp_code}
\end{center}
\end{figure}

We tested two methods using OpenMP.
First, at each iteration we run, in sequence, five parallel for-loops, using {\fontfamily{pcr}\selectfont \#pragma omp parallel for}.
Each parallel for-loop updates all variables of the same kind (e.g. all the $x$ variables).
Second, we create a parallel section, using {\fontfamily{pcr}\selectfont \#pragma omp parallel}, in which each thread processes all updates across multiple iterations (this approach requires {\fontfamily{pcr}\selectfont \#pragma omp barrier} to synchronize threads between updates types, such as $x$ to $m$).
Figure \ref{fig:openmp_code} gives more detail.
We found the first approach to be substantially faster and we report its speed-up results in the numerical section.

The user can change {\fontfamily{pcr}\selectfont N\_threads} to control the number of cores used.
To solve a problem using one CPU-core the user can set this value to $1$ or replace OpenMP update calls with serial versions (e.g. {\fontfamily{pcr}\selectfont updateXOpenMP} to {\fontfamily{pcr}\selectfont updateX}).

\section{Related work}
\label{sec:related_work}
\noindent
Before we show our numerical results we present a review of related work.
The focus of this paper is the ADMM, and schemes for [universal] acceleration using multiple-cores.
However, it is useful to begin by showing that there are very few general-purpose optimization engines that can automatically exploit multi-core computing (especially GPUs), which indicates the importance of our line of research.
In addition, reporting the range of speedups that other work achieves helps contextualize our results: in particular, even on a GPU with thousands of cores, one should expect much less than $1000\times$ acceleration.

\begin{figure}[t!]
\begin{center}
\includegraphics[trim = 0mm 0mm -120mm 0mm, clip, scale=0.3]{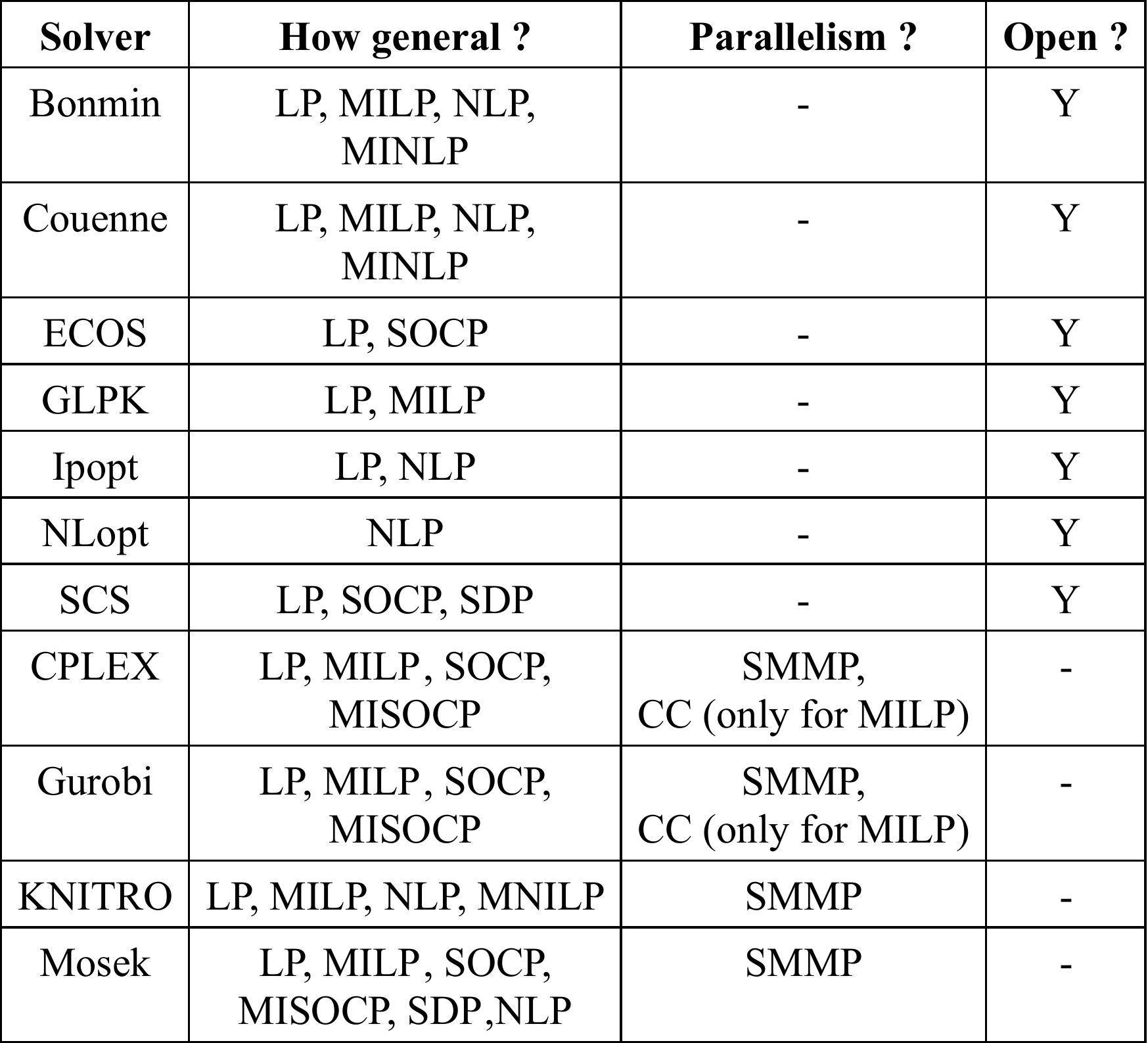}
\put(-100,130){\tiny LP - linear programming}
\put(-100,120){\tiny NLP - non-linear programming}
\put(-100,110){\tiny MILP - mixed integer LP}
\put(-100,100){\tiny MINLP - mixed integer NLP}
\put(-100,90){\tiny SOCP - second-order cone programming}
\put(-100,80){\tiny MISOCP - mixed-integer SOCP}
\put(-100,70){\tiny SDP - semi-definite programming}
\put(-100,20){\tiny SMMP - shared-memory multi-processor}
\put(-100,10){\tiny CC - computer cluster}
\caption{State-of-the-art optimization solvers.}
\vspace{-0.2cm}
\label{fig:table_of_solvers_and_parallelism}
\end{center}
\end{figure}

Figure \ref{fig:table_of_solvers_and_parallelism} summarizes the picture for solvers that just use
CPUs: most open-source solvers cannot exploit parallelism; commercial solvers allow parallelism in shared-memory multi-processor frameworks for special classes of optimization problems, and only a few can directly use multiple independent machines. In the context of using multiple CPU-cores there is also the recent SNAPVX \cite{hallac2015snapvx}, which is also a factor-graph based implementation of the ADMM.
It is written in Python and hence much slower than \textit{parADMM}.
It forces users to use CVXPY \cite{diamond2015cvxpy}, which introduces additional delays and restricts users to convex problems only, while {\it parADMM} supports non-convex problems.
In addition, it can only solve problems that decompose into a very specific factor-graph (function nodes of degree two) and does not support GPUs.

%

There are a number of GPU-accelerated, population-based, general-purpose optimization heuristics that come with no convergence guarantees.
These exploit the GPU by assigning different parts of the exploring population to different GPU-cores.
For example, \cite{mussi2011evaluation} studied particle-swarm optimization algorithms and reported speedups of $\sim 30\times$ with a 1.86GHz Core i2 vs. an Asus GeForce EN8800GT.
\cite{schulz2013gpu} reported on several GPU-accelerated optimization solvers for discrete optimization; these are either general-purpose heuristics or problem-specific (e.g. routing).

There is work on solving linear programs using GPUs to speedup linear-algebra computations, all of which come with guarantees.
The work of \cite{bieling2010efficient} focuses on speeding up the Revised Simplex method and achieves a speed-up of $18\times$ with a 3.0GHz Core i2 vs. an NVIDIA GeForce 9600GT.
Later work \cite{lalami2011multi} reports a speedup of about $24\times$ on accelerating the Simplex method, but uses two NVIDIA Tesla C2050 GPUs against a single core of a 2.66GHz Xeon E5640.
\cite{smith2012gpu} accelerates the matrix-free interior point method for linear programs and achieved $4\times$ speed-up for a 2.4GHz AMD Opteron 2378 core vs. an NVIDIA Tesla C2070.

There is also an increasing set of libraries/tools that make the use of multiple CPU cores, GPUs, or multiple computers easy: OpenCL, OpenACC, OpenHMPP, OpenMP, C++ AMP, Open-MPI, Hadoop, Apache Spark, etc.
In particular, the first three support GPUs and so, in principle, some of the existing open-source solvers, can be (at least in part) re-written using these tools for GPUs; however, we are not aware of such work.
There are numerous examples using such tools to code solvers for specific problems that can exploit distributed computation and speed time-to-solution;
however, many of the algorithms used internally for the open-source, general-purpose solvers do not facilitate automatic GPU exploitation, which is likely why they are not re-written.

It is important to make explicit the novelty of our work compared to related ADMM-based projects.
There are a few projects that use the ADMM to exploit GPU parallelism; however, these are mostly problem specific.
For example, \cite{cui2011distributed} use a consensus formulation of the ADMM to perform image recovery for Positron Emission Tomography and for Computed Tomography.
They write a problem-specific GPU-ADMM solver and obtain a $200\times$ speedup on a small cluster of two computers, each with two GPUs, as compared to their own single-thread implementation.
The authors also evaluate the speed of a GPU-distributed-gradient-descent solver and report that this method is slower than using GPU-ADMM.
\cite{bilen2012high} looked at an inverse problem in Magnetic Resonance Imaging and proposed an ADMM-based solution to this specific problem.
Using a GPU and the JACKET 2.0 toolbox for MATLAB, they speedup their solver up to $10\times$ in some cases.
\cite{bhaskar2011admm} also use JACKET to accelerate the algebra calculations in an ADMM-based solution to the sparse coding problem and achieve a speedup of $8\times$ over a serial version of their code.
\cite{de2011alternating} implemented the ADMM in a GPU for deblurring images and obtain a speed-up of $25\times$ compared to another commonly used serial algorithm.
\cite{miksikdistributed} solved the problem of inference in large-scale random fields using a GPU-accelerated ADMM implementation: for this specific problem they achieve a speedup of $100\times$ over a serial ADMM implementation and a speedup of $5\times$ against using a GPU-accelerated classic dual decomposition method.

We note that there do exist some close-to-general-purpose, open-source, ADMM-based parallel optimization frameworks.
\cite{lawson2014alternating} proposes a framework that uses Spark and computer clusters to accelerate the ADMM.
This framework is restricted to consensus formulations of the ADMM (the factor-graph is star shaped) and does not support the use of GPUs.
\cite{fougner2014pogs} proposes a framework that can use the CUDA Thrust library to accelerate the computation of the POs and can also use a GPU to accelerate matrix computations;
however this framework only supports decomposing the original problem into two sub-problems using functions $f$ and $g$ that are fully separable, restricting the objective function to a compositions of a small number of convex functions from a predefined library, and does not support the same level of automatic parallelism as ours. The author reports a speed-up of $13-30\times$ vs. a single CPU core to solve a Lasso problem.

\section{Numerical results}
\noindent
In this section we report the speedup for our automatic-parallelization ADMM scheme. 
In the context of GPUs, as far as we know, we are the first to systematically evaluate this scheme on very different problems using the same tool.

The GPU speedups compare the runtime of the ADMM on a single core of an AMD Opteron Abu Dhabi 6300 at 2.8GHz with the runtime of the ADMM on a NVIDIA Tesla K40 GPU for the same number of iterations. 
The multi CPU-core speedups compare the runtime on a different number of cores (up to $32$) of an AMD Opteron Abu Dhabi 6300 at 2.8GHz. 
The machine that hosts the GPU and the $32$ AMD cores has $128$GB of memory and runs Ubuntu Linux.
All the code is in C, CUDA, and OpenMP and is compiled using \texttt{gcc} and \texttt{nvcc} on Ubuntu using the choice of optimization flags that gives the best performance. 
When using OpenMP we use the linux command \texttt{nice -n -15} to give the processes high priority.
Our code is open source (\url{https://github.com/parADMM/engine}).

We now study three problems from three different and important fields: combinatorial optimization, optimal control, and machine learning.

\subsection{Combinatorial optimization}
\noindent
Packing problems are essential
to fields like
condensed matter physics and coding theory.
The classical packing problem inquires as to the
maximum number of 2D disks of radius $R$ that can be
positioned entirely within a 2D unit square, but there are many variants
of this problem and packing is very much
an active area of research. For example, a complete formal proof
of Kepler's conjecture, that heavily relies on computers,
was only published recently \cite{hales2015formal}.
Many conjectures remain open and several problems can benefit
if we can use a computer to produce or validate different large
packing configurations fast.

Here we show how to
formulate an NP-hard packing problem as an optimization problem --
we then heuristically solve it using the ADMM and show how much our tool
accelerates the ADMM. This is a good idea since the authors in \cite{Derbinsky2013AnImp,derbinsky2014scalable} show that using the ADMM to solve packing problems produces record-breaking packing densities.
Our specific packing problem is the following:
given $N$ non-overlaying disks with center $c_i$ and radius $r_i$
inside a triangle $\mathcal{T}$, what is the largest area they can cover?
This problem is related to the open problem of extending Malfati's circle conjecture to an arbitrary number of disks \cite{andreatta2011problem}.
%
%

\begin{figure}[t!]
\begin{center}
\includegraphics[trim=0cm 0cm 0cm 0cm, clip=true,scale=0.25]{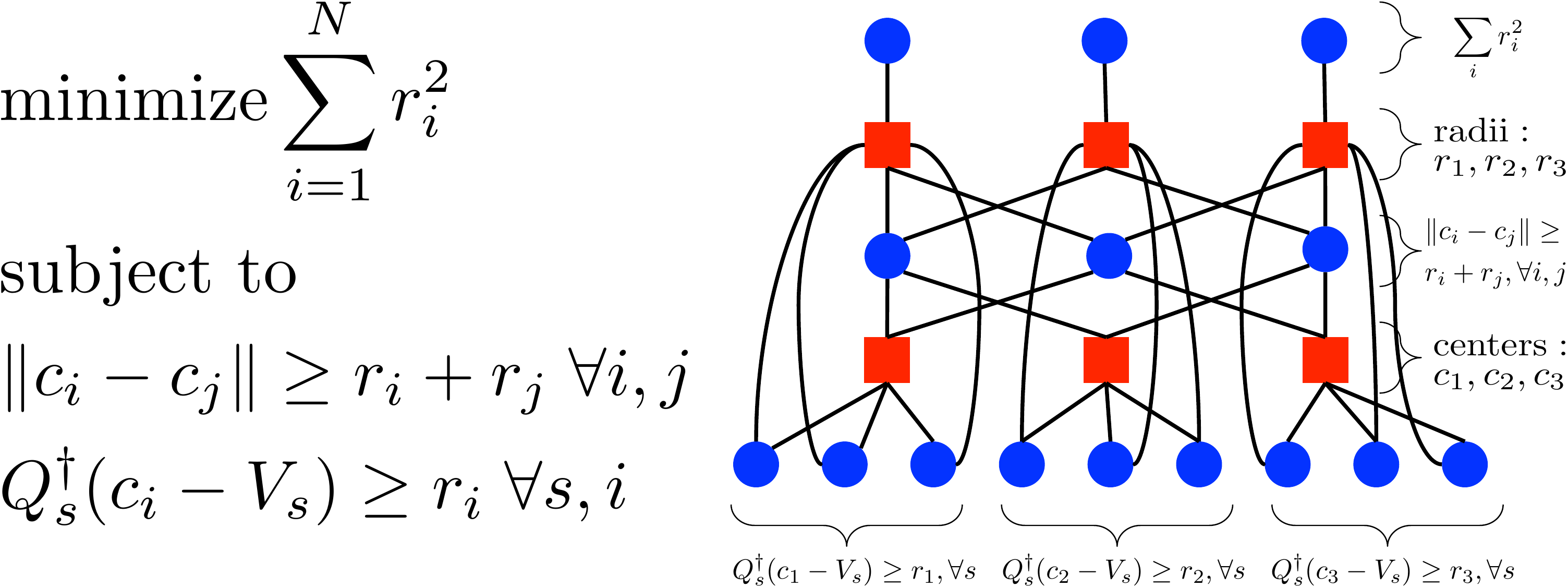}
\caption{Optimization problem: cover a triangle with disks.\vspace{-0.3cm}}
\label{fig:circle_packing_equations_and_graph}
\end{center}
\end{figure}

Figure \ref{fig:circle_packing_equations_and_graph} shows the formulation
we use and the
factor-graph decomposition used for the ADMM exemplified for $N=3$.
To impose that each disk must lie within the triangle $\mathcal{T}$ we
impose that each disk is inside three half-planes, $s = 1,2,3$, specified by
their normal direction, $Q_s$, and a point in the plane, $V_s$.

According
to this formulation, the ADMM factor-graph for $N$ circles and a box formed
by the intersection of $S$ half-planes has $2N^2 - N + 2NS$ edges, $2N$ variable nodes and $0.5N(N-1) + N + NS$ function nodes. The number of
elements in the factor-graph grows quadratically with $N$ and all proximal operators have closed-form solutions, a setting in which \textit{parADMM} is well suited to accelerate the ADMM.

\begin{figure}[b!]
\begin{center}
\includegraphics[trim=0cm 0cm 0cm 0cm, clip=true,scale=0.25]{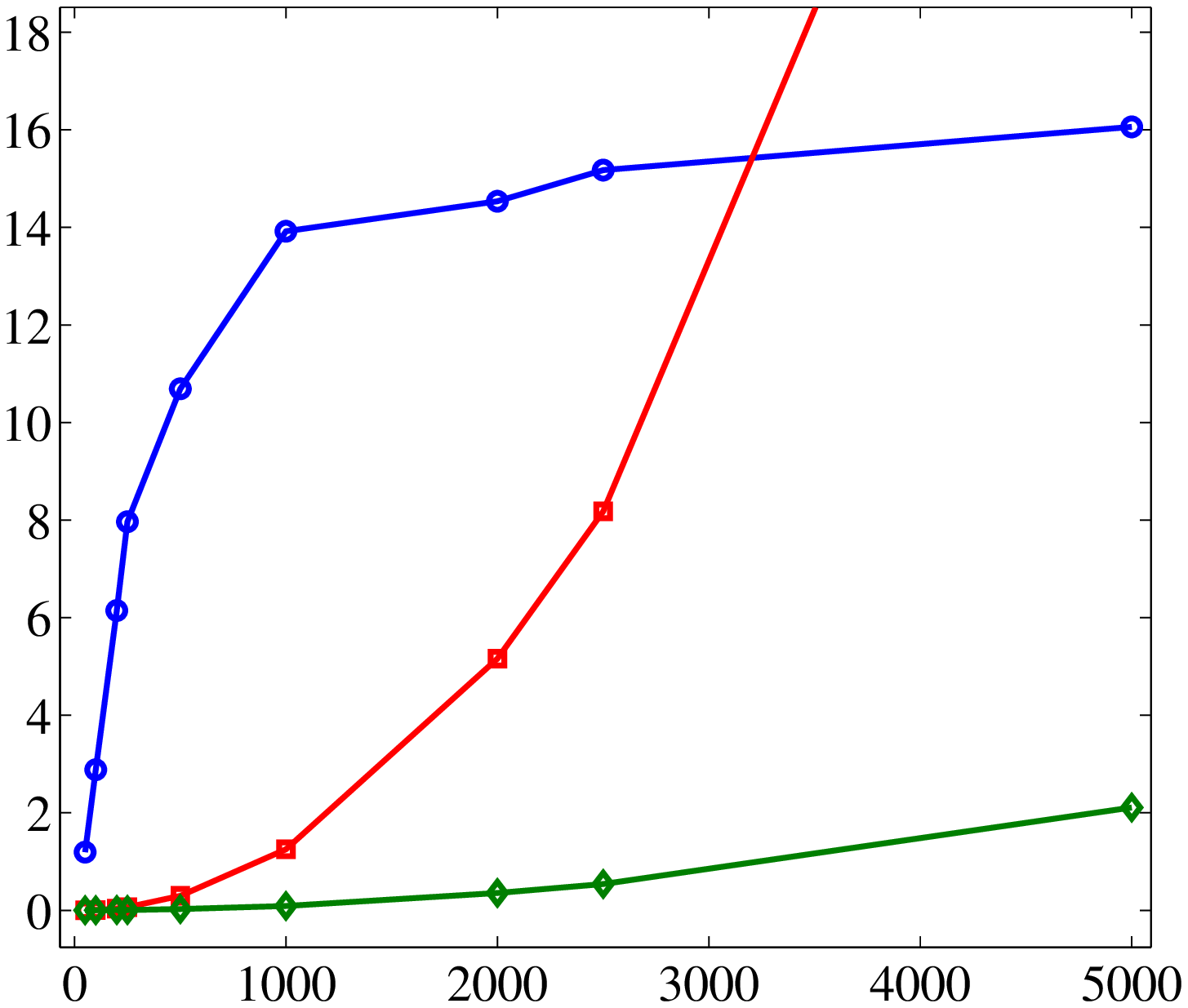}
\put(-80,-02){\tiny Number of circles $N$}
\put(-80,70){\tiny Speedup}
\put(-56,50){\tiny CPU-time}
\put(-40,24){\tiny GPU-time}
\put(-130,33){\tiny \rotatebox{90}{Combined speedup}}
\put(-120,25){\tiny \rotatebox{90}{Time for $10$ iterations (s)}}
\includegraphics[trim=0cm 0cm 0cm 0cm, clip=true,scale=0.267]{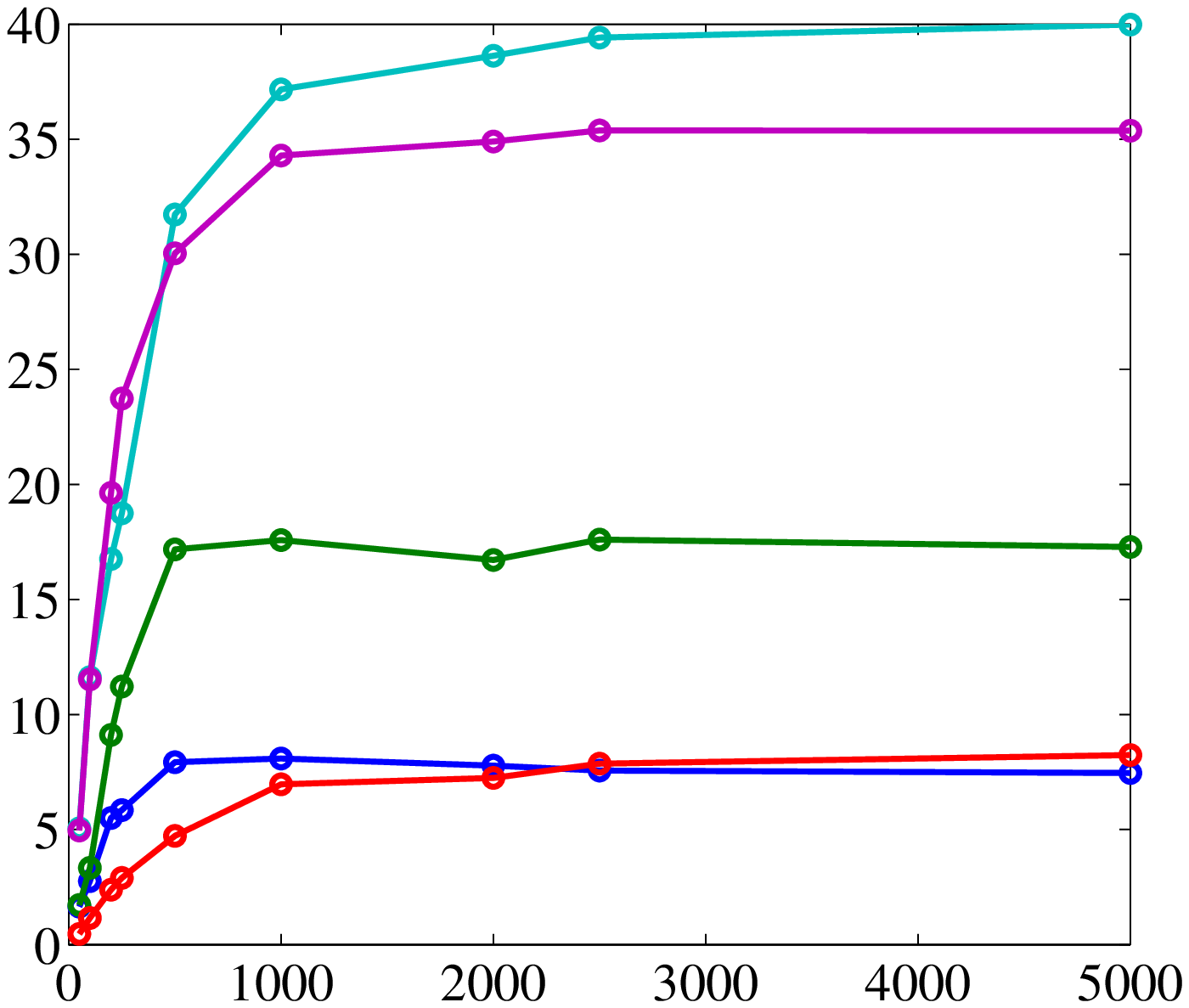}
\put(-80,-02){\tiny Number of circles $N$}
\put(-40,20){\tiny x-update}
\put(-40,30){\tiny z-update}
\put(-40,49){\tiny m-update}
\put(-40,90){\tiny u-update}
\put(-40,81){\tiny n-update}
\put(-125,33){\tiny \rotatebox{90}{Individual speedups}}
\caption{GPU vs. CPU in packing. (Left) Combined updates; (Right) Individual updates.
\vspace{-0.3cm}}
\label{fig:speed_up_circle_packing}
\end{center}
\end{figure}

Figure \ref{fig:speed_up_circle_packing} shows that we can get more than $16\times$ speedup using a GPU vs. single CPU-core for large $N$.
It also shows that the time per iteration grows linearly with the number of elements in the
factor-graph, which we know grows quadratically in $N$. The slowest updates
are the $x$ and $z$ updates that (for $N=5000$) take  $31\% + 40\% = 71\%$ of the time respectively.
These are also the hardest steps to speedup as
Figure \ref{fig:speed_up_circle_packing}-right shows.
For most of the points in the plots above we find that using $32$ threads per block gives the best performance. For example, the speedup in the $x$-update for $N = 5000$ circles takes the values $5.6, 5.6, 5.8,5.8,5.8,7.4,5.5,3.5,2.0,2.0,3.6$ for $ntb = 1,2,4,8,16,...,512$ respectively.

We note that the time to copy the result from the GPU back to the CPU,
for example for the purpose of checking stopping criteria, is negligible.
For $N = 5000$ it takes only $0.3ms$ to copy $z$ from the GPU to the CPU.
The time it takes to create the factor graph and copy it to the GPU can take
some time, up to $450sec$ for $N=5000$. However, this time is still
negligible compared to the time to run enough iterations for convergence
(\textgreater hundreds of thousands for $5000$ circles). Also, once formed and copied to
the GPU the graph can be reused for different instances of similar problems.

\begin{figure}[t!]
\begin{center}
\includegraphics[trim=0cm 0cm 0cm 0cm, clip=true,scale=0.235]{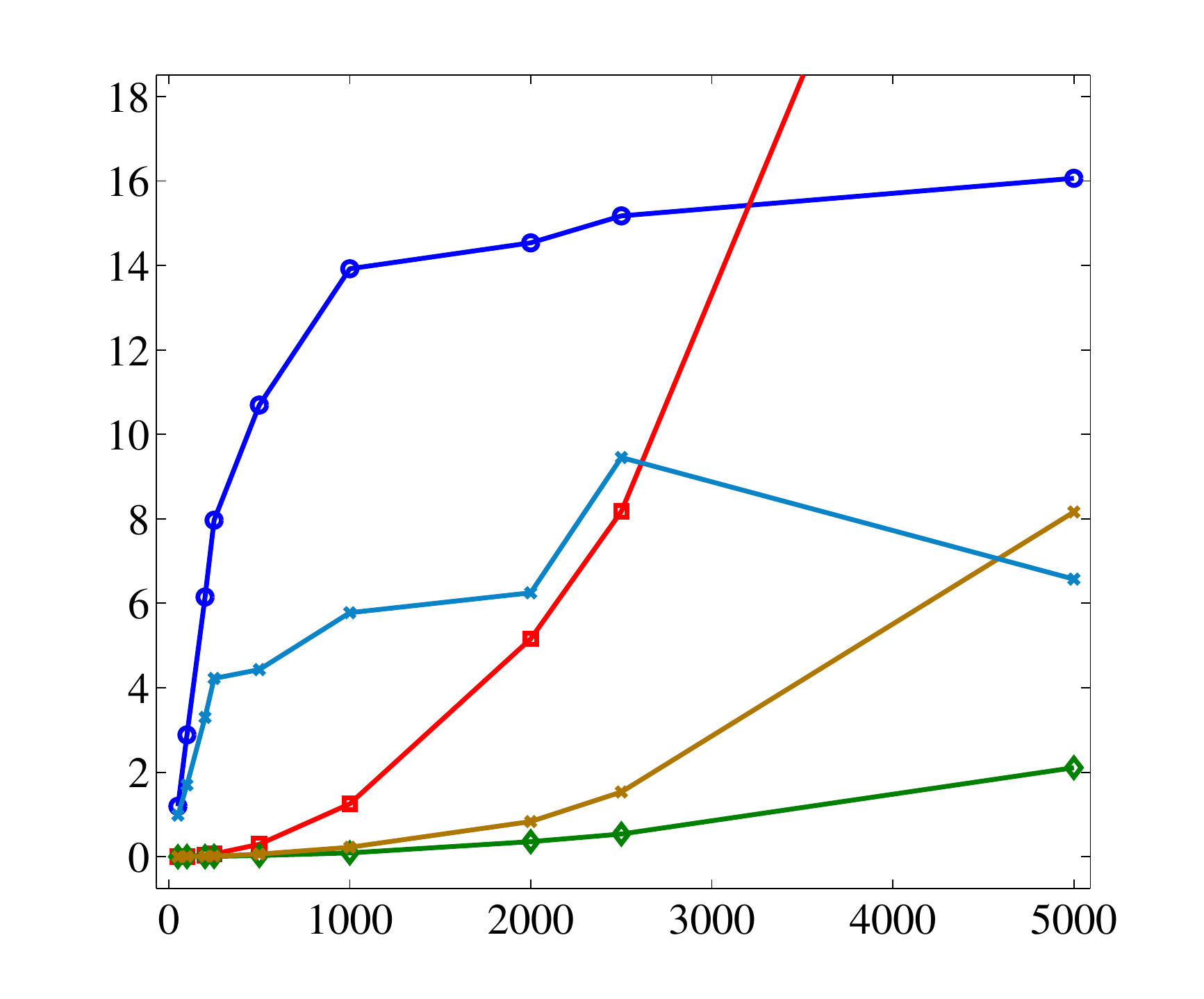}
\put(-80,-02){\tiny Number of circles $N$}
\put(-97,82){\tiny GPU speedup}
\put(-100,42){\tiny Multi CPU-core}
\put(-100,36){\tiny speedup}
\put(-56,65){\tiny CPU time}
\put(-40,16){\tiny GPU time}
\put(-60,24){\tiny Multi CPU-core time}
\put(-130,43){\tiny \rotatebox{90}{Speedup}}
\put(-120,20){\tiny \rotatebox{90}{Time for $10$ iterations (s)}}
\includegraphics[trim=0cm 0cm 0cm 0cm, clip=true,scale=0.24]{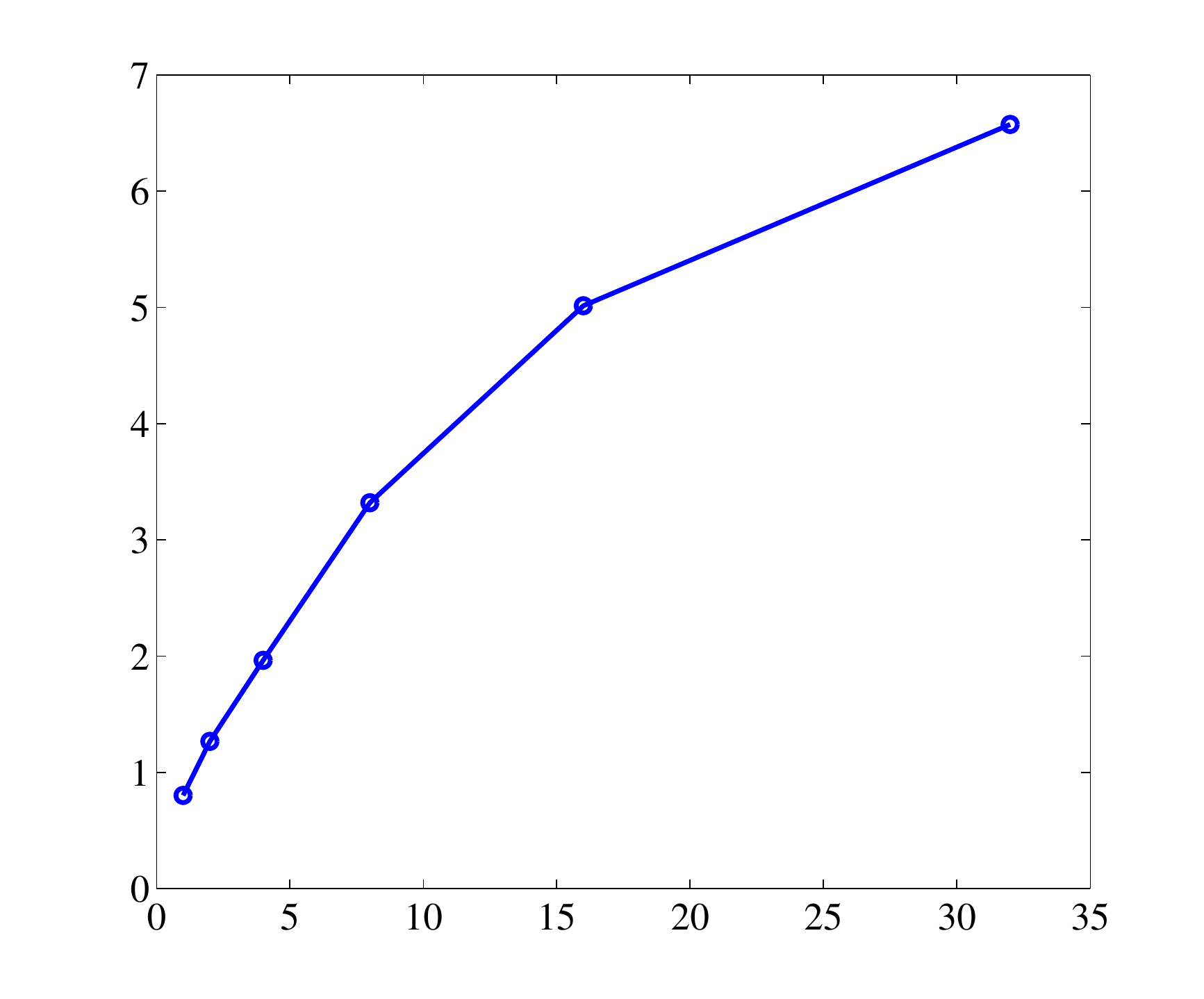}
\put(-78,-02){\tiny Number of CPU cores}
\put(-120,27){\tiny \rotatebox{90}{Multi CPU-core speedup}}
%
%
%
\caption{Multiple CPU vs. single CPU in circle packing. (Left) Combined speedup. (Right) Speedup vs. number of cores for $N = 5000$.
\vspace{-0.2cm}}
\label{fig:open_MP_circle_packing_times}
\end{center}
\end{figure}
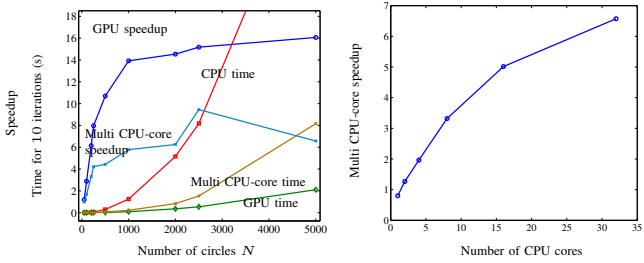

Figure \ref{fig:open_MP_circle_packing_times}-left shows that
we can get up to $9\times$ speedup using $32$ CPU-cores, substantially less than the $16\times$ with a GPU.
In addition, the $9\times$ speedup holds only for special values
of $N$ (around $2500$) and drops to $6\times$ for larger problems.
In our experiments, the combined $x$ and $z$ updates now only take $18\% + 11\% = 29\%$ of the time per iteration respectively (for $N = 5000$) and the CPU-cores, unlike the GPU-cores, produce similar speedups regardless of them updating $x, m, z, u$ or $n$.
Figure \ref{fig:open_MP_circle_packing_times}-right shows
that for large problems, the speedup starts saturating
with more cores.

%

%
%
%
%
%
%

As a reference we note that, for example, on a single core and for $500$ circles,
the time per iteration of our tool is more than $4\times$ faster than the tool used
by \cite{Derbinsky2013AnImp,derbinsky2014scalable}.

\subsection{Optimal control}
\noindent
In Model Predictive Control (MPC) we predict how the trajectory of a system evolves
under different inputs and with this knowledge we optimally
lead the system to a desired state.
MPC was first used to control chemical processes but since then it has been used
in many other applications (c.f. \cite{garcia1989model} for a good survey). MPC can be useful both offline (real-time performance does not matter) and online (real-time performance matters).

Here we test our ADMM scheme for both GPU and multiple CPU-cores when solving an MPC problem for a discrete-time
linear system $q(t+1) - q(t) = A q(t) + B u(t)$, where $q(t)$ is the state of the system at time $t$ and $u(t)$ is the input to the system at time $t$.
Note that we are not the first to use a GPU to solve MPC problems, see for example
\cite{gade2012mpc}. However, the latter work used a linear-programming interior point method that required writing problem specific CUDA code while we only wrote serial code for each PO.

\begin{figure}[t!]
\begin{center}
\includegraphics[trim=0cm 0cm 0cm 0cm, clip=true,scale=0.3]{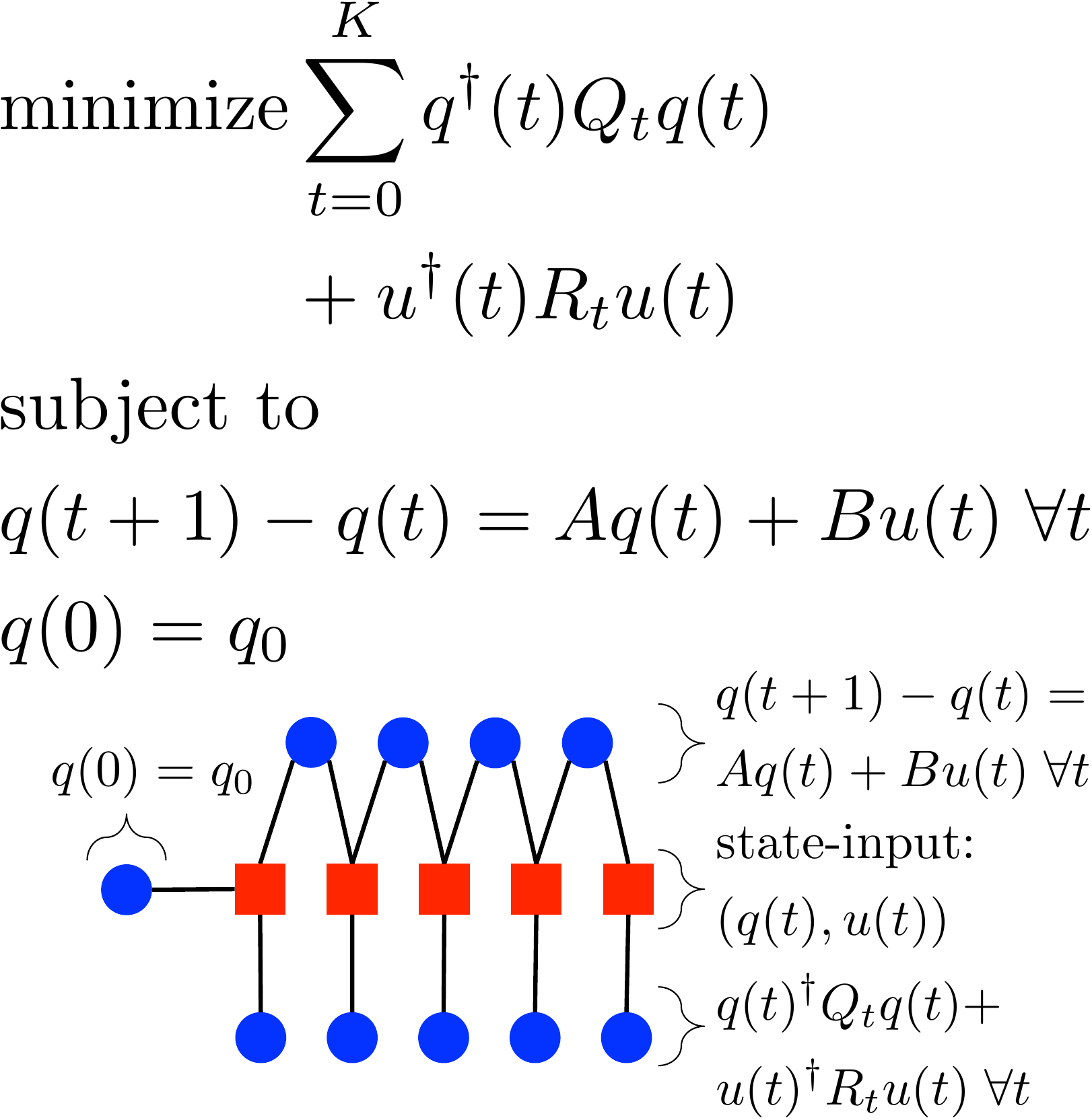}
\caption{MPC formulation and factor-graph for $K = 4$.}
\label{fig:MPC_formulation}
\end{center}
\end{figure}

There
are many different MPC formulations including, but not limited to, robust MPC, non-linear MPC and closed-loop MPC. In Figure \ref{fig:MPC_formulation} we show
the specific formulation we use and its corresponding factor-graph.
In the factor-graph, each variable node is associated to one $q(t)$ and one $u(t)$.
The variable $K$
is the prediction horizon and we vary it between $200$ and $10^5$
to test how the speedup depends on the size of the problem.
Note that the number of elements in the factor-graph grows linearly with $K$.
In our tests we have $A \in \mathbb{R}^{4 \times 4}$ and $B  \in \mathbb{R}^{4 \times 1}$ and both are obtained from linearizing (around equilibrium) and sampling (every $40$ ms) a continuous time inverted-pendulum system.
The matrices $Q$ and $R$ are a design choice and, for simplicity sake, we make all $\{Q_t\}$ equal and all $\{R_t\}$ equal and we make each of them diagonal. Finally, $q_0$ is the known state of the system at the instant from which we predict its future behavior.

\begin{figure}[b!]
\begin{center}
\includegraphics[trim=0cm 0cm 0cm 0cm, clip=true,scale=0.3]{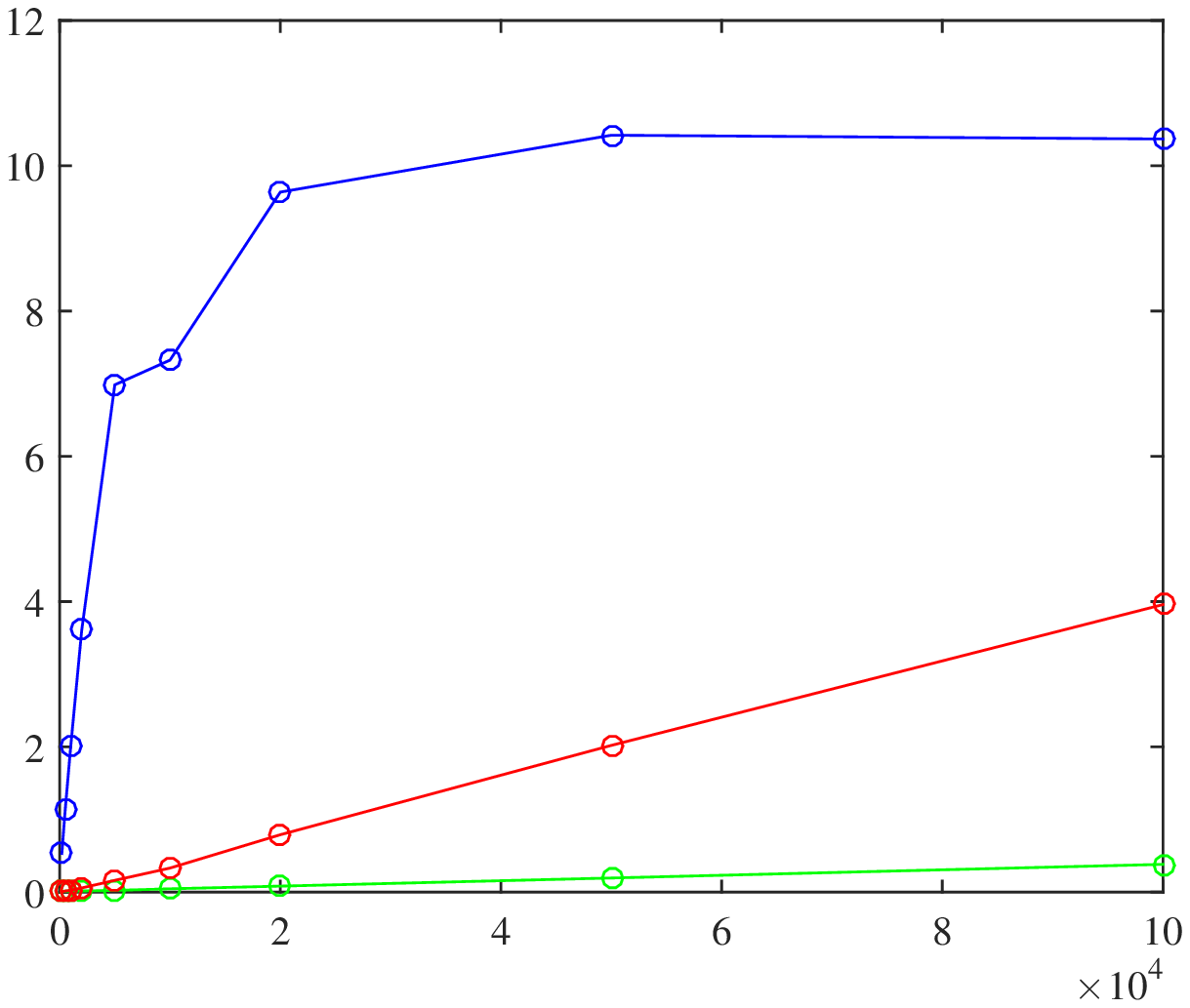}
\put(-90,-01){\tiny Prediction Horizon $K$}
\put(-85,25){\tiny CPU time}
\put(-60,80){\tiny Speedup}
\put(-40,15){\tiny GPU time}
\put(-130,43){\tiny \rotatebox{90}{Speedup}}
\put(-122,20){\tiny \rotatebox{90}{Time for $100$ iterations (s)}}
\includegraphics[trim=0cm 0cm 0cm 0cm, clip=true,scale=0.3]{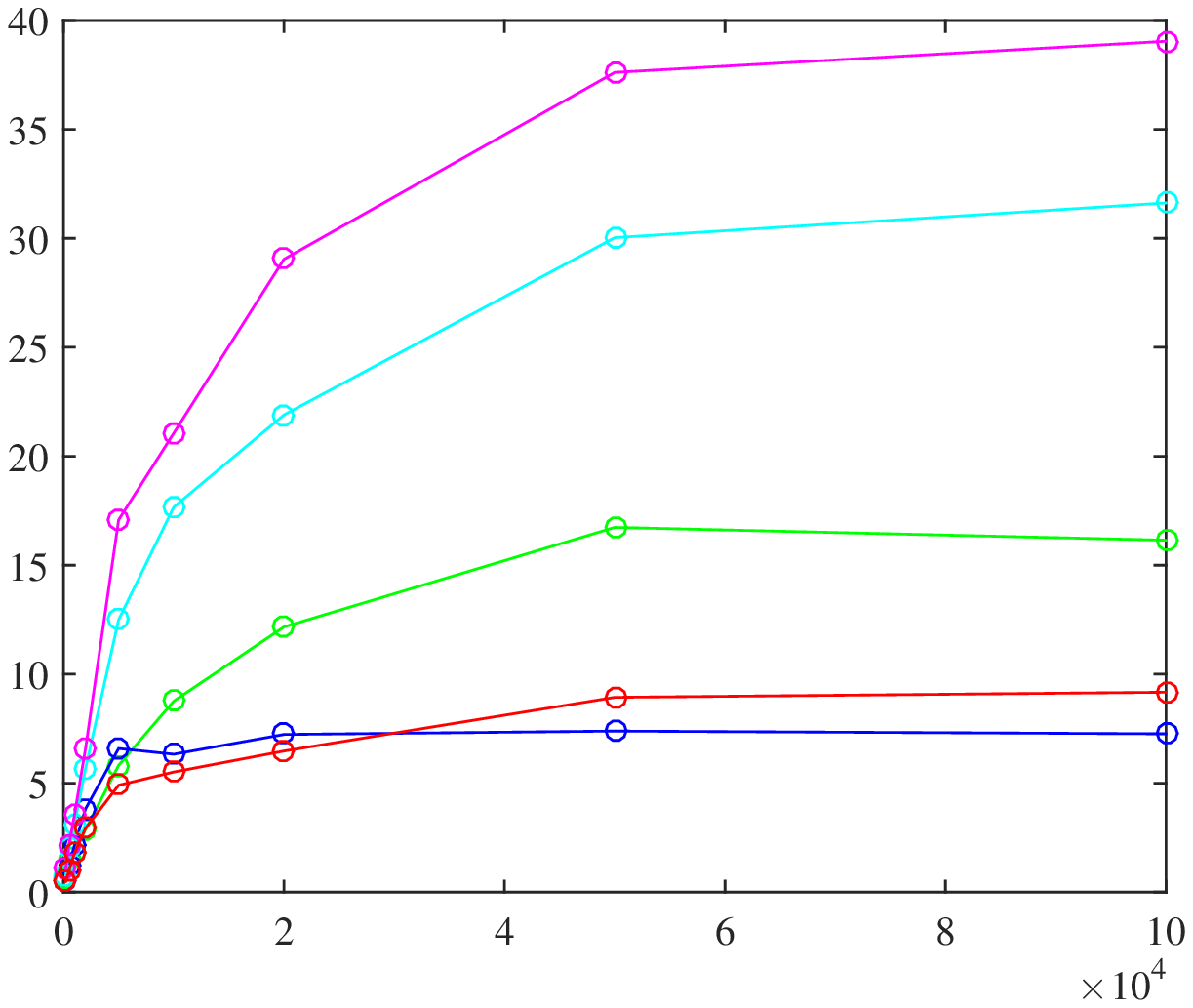}
\put(-90,-01){\tiny Prediction Horizon $K$}
\put(-40,20){\tiny x-update}
\put(-40,30){\tiny z-update}
\put(-40,43){\tiny m-update}
\put(-40,81){\tiny n-update}
\put(-40,66){\tiny u-update}
\put(-125,43){\tiny \rotatebox{90}{Speedup}}
\caption{GPU speedup for MPC. (Left) Combined updates; (Right) Individual updates.\vspace{-0.5cm}}
\label{fig:speed_up_MPC}
\end{center}
\end{figure}

Figure \ref{fig:speed_up_MPC}-left shows that we can
get up to $10\times$ speedup for large problems. As expected,
the time per iteration grows linearly with the number of elements in the factor-graph, which we know grows linearly with $K$. Like in circle packing
the $x$ and $z$ updates are the slowest updates and take
$59\% + 21\% = 80\%$ of the time per iteration respectively ($K = 10^5$)
and also the hardest to speedup as Figure \ref{fig:speed_up_MPC}-right shows.
For MPC we again find that using $ntb = 32$ threads per block gives the best performance. The exception is the z-update for which we find that using a smaller $ntb$ gives better performance.
More specifically, for $K = 200, 10^3, 10^4, 5\times10^4, 10^5$ we find that the optimal $ntb$ in the z-update are $2, 8, 16, 16, 16$ respectively.

The time to copy the final result from the GPU to the CPU
is negligible, about $3ms$ for $K = 10^5$. The time to copy the factor-graph
from the CPU to the GPU can take up to $13$ seconds for $K = 10^5$, which
is negligible also compared to the number of iterations until convergence (more
that several thousands for $K = 10^5$) and a random initialization of the ADMM.
In addition, in the context of solving a
problem in real-time, we only need to create and move the graph to the
GPU once. In each cycle of our feedback controller we only need to update
the value in the GPU of the current state of the system, which can be done almost
instantaneously and then run a few more ADMM iterations on the factor-graph already on the GPU starting from the ADMM solution of the previous cycle.

\begin{figure}[t!]
\begin{center}
\includegraphics[trim=0cm 0cm 0cm 0cm, clip=true,scale=0.3]{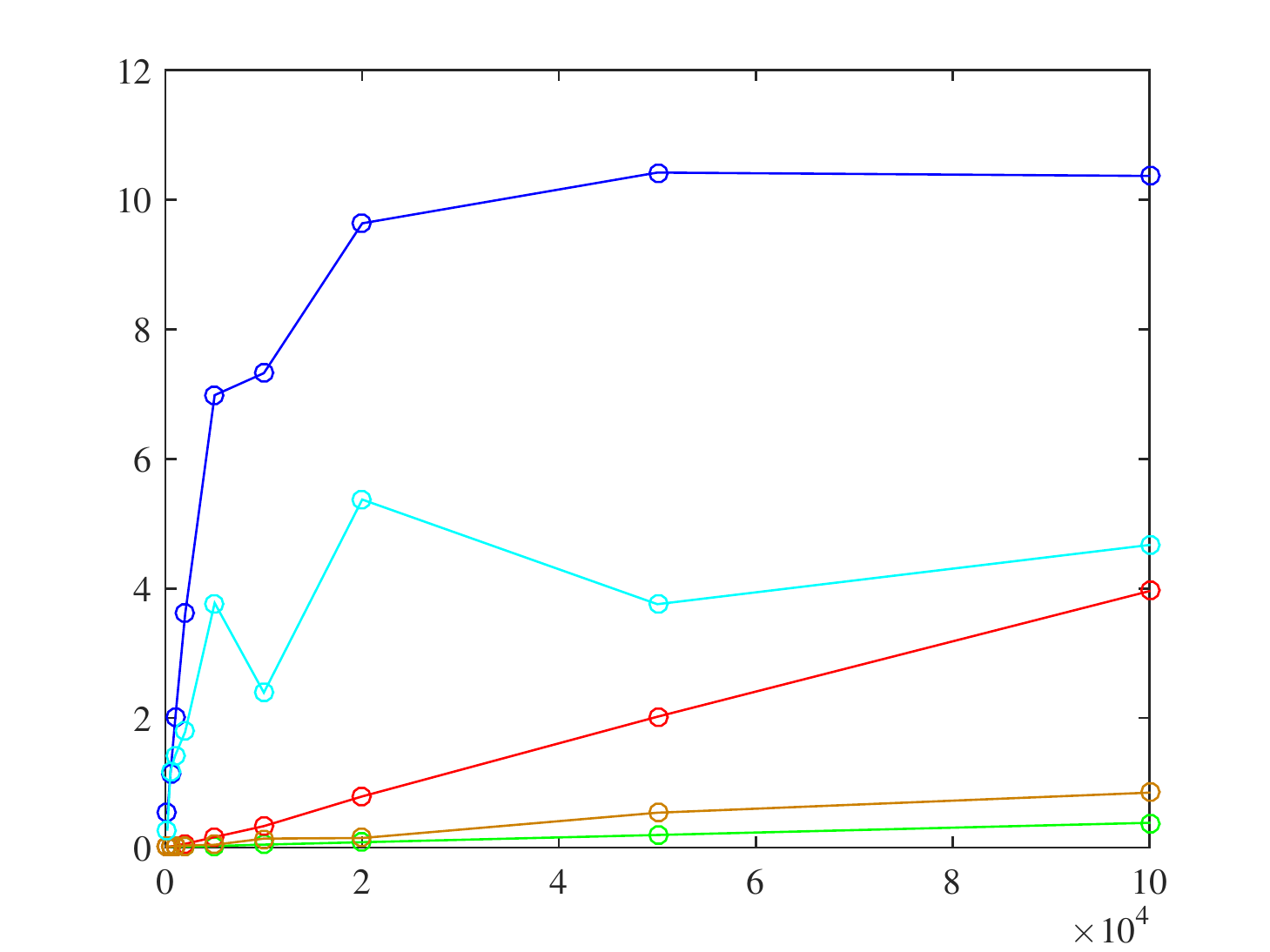}
\put(-90,-01){\tiny Prediction Horizon $K$}
\put(-85,25){\tiny CPU time}
\put(-85,43){\tiny Multi CPU-cores speedup}
\put(-61,80){\tiny GPU speedup}
\put(-60,17){\tiny Multi CPU-cores time}
\put(-37,12){\tiny GPU time}
\put(-130,43){\tiny \rotatebox{90}{Speedup}}
\put(-122,20){\tiny \rotatebox{90}{Time for $100$ iterations (s)}}
\includegraphics[trim=0cm 0cm 0cm 0cm, clip=true,scale=0.3]{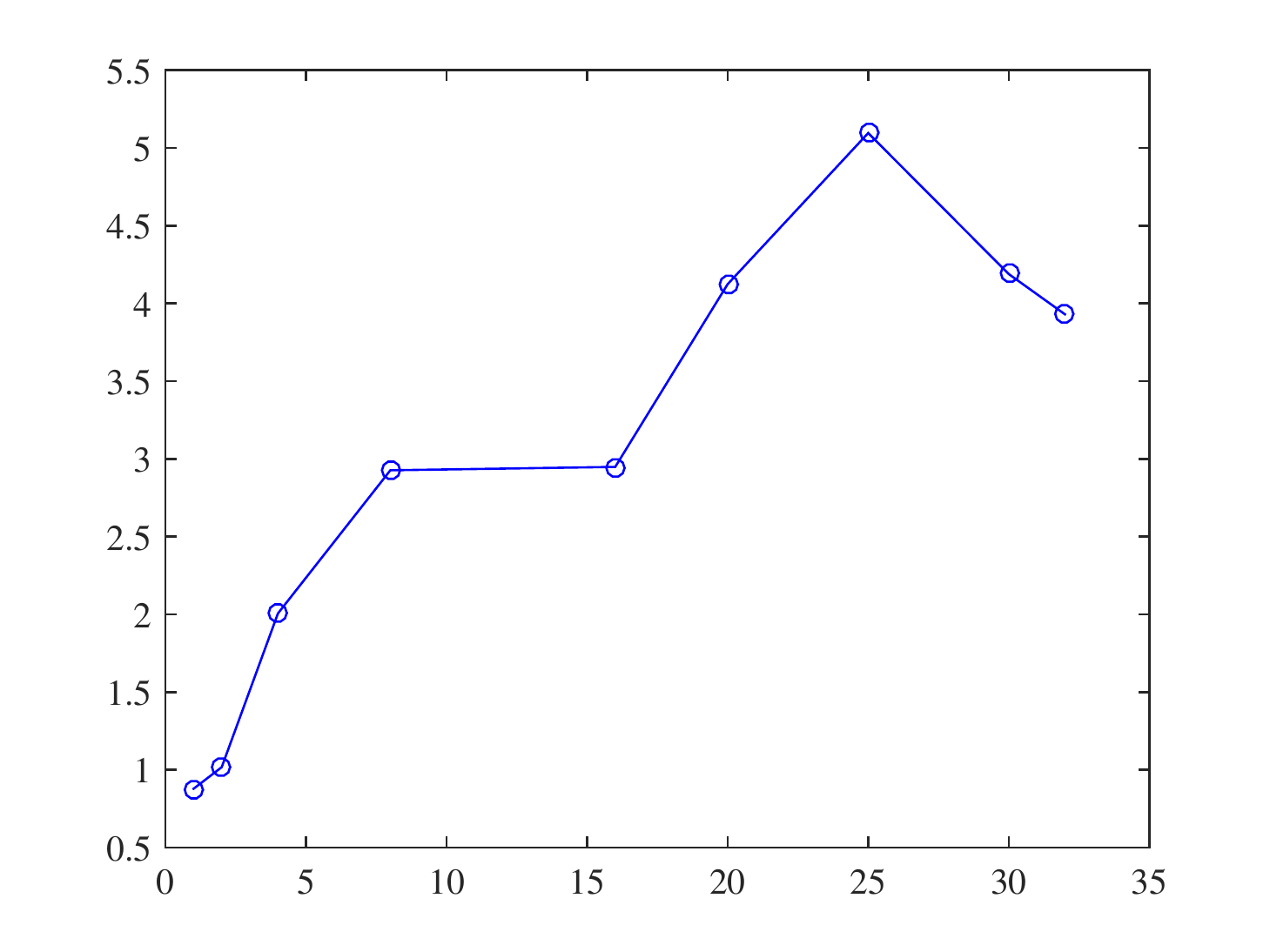}
\put(-85,-01){\tiny Number of CPU cores}
\put(-124,27){\tiny \rotatebox{90}{Multi CPU-core speedup}}
\caption{Multi CPU-cores speedup for MPC. (Left) Combined updates (Right) Speedup vs \# of threads for $K$ (prediction horizon) $=10\times 10^4$.\vspace{-0.4cm}}
\label{fig:MPC_OpenMP}
\end{center}
\end{figure}
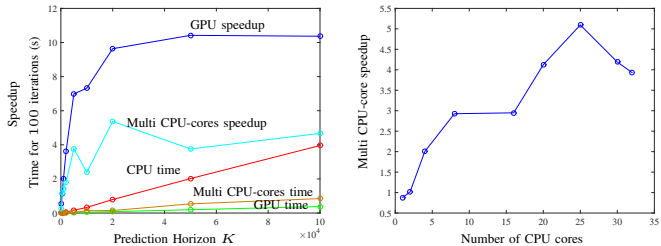

In Figure \ref{fig:MPC_OpenMP}-left we report the test of our fine-grained parallelism using $25$ CPU cores. We use $25$ cores since this seems to produce the highest speedup. In fact, Figure \ref{fig:MPC_OpenMP}-right shows that for large problems, as we add more cores, the performance actually gets hurt. The best speedup is about $5\times$ and, just like
for circle packing using multiple CPU-cores, varies a bit irregularly with the problem size. When using multiple cores, the slowest updates are the
$m$, $u$ and $n$ updates which take $25\% + 19\% + 16\% = 60\%$ of
the time per iteration respectively (for $K = 10^5$).

\subsection{Machine learning}
\noindent
Support Vector Machines (SVM) are used successfully in many real world  problems, e.g. cancer diagnosis \cite{furey2000}. In this section we work with the soft margin SVM but there are many other different variants of SVM. In this formulation, given a data set of $N$ vectors $\{x_i\}^N_{i=1}$ with labels $\{y_i\}^N_{i=1}$ taking values in $\{-1,+1\}$
the objective is to find the slab $(w,b)$ that separates the $+1$ vectors from the $-1$ vectors as well as possible. The quality of the separation is measured by the norm of $w$ and by $N$ non-negative slack variables $\{\xi_i \}^N_{i=1}$.

\begin{figure}[t!]
\begin{center}
\includegraphics[trim=0cm 0cm 0cm 0cm, clip=true,scale=0.27]{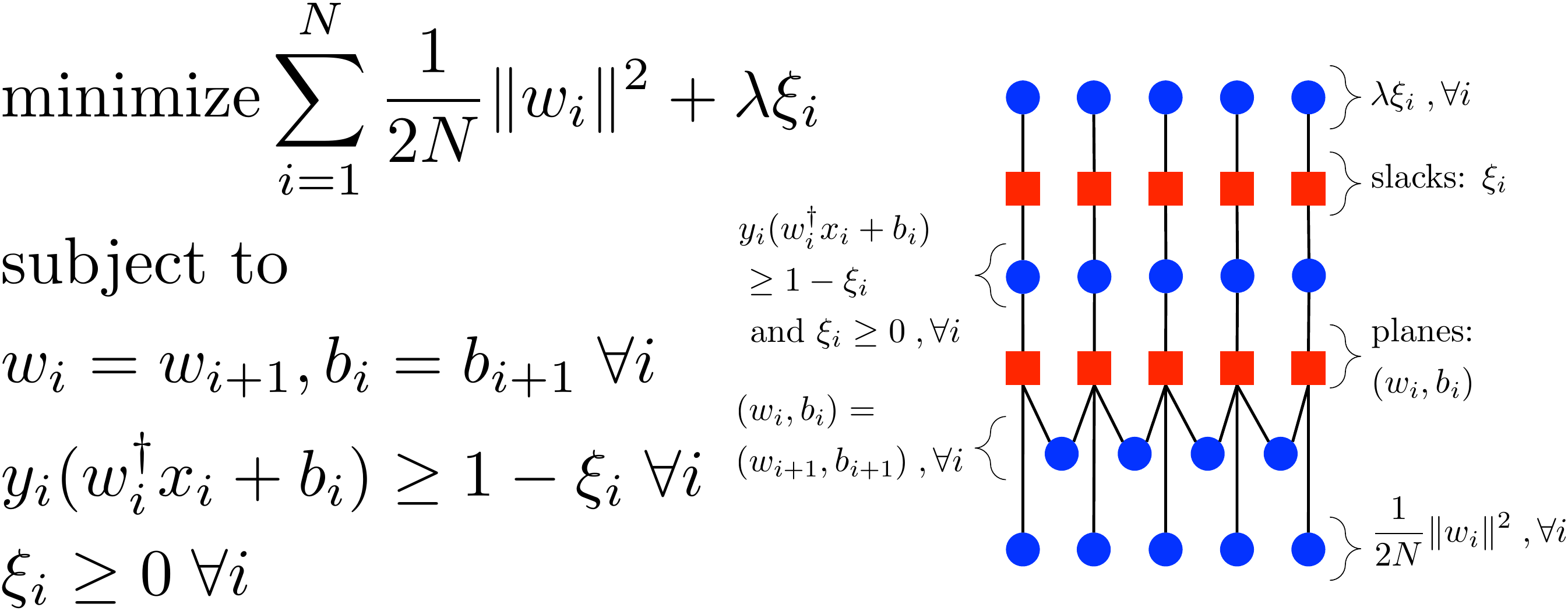}
\caption{Optimization problem for soft-margin SVM.}
\label{fig:SVM_formulation_and_graph}
\end{center}
\end{figure}

Figure \ref{fig:SVM_formulation_and_graph} gives the details of our formulation and factor-graph decomposition.
Notice that we create multiple copies of the $w$ variable and break the
term in the objective function associated with $w$ into $N$ equal parts.
This makes the distribution of the number of edges-per-node in the
factor-graph more equilibrated, which  in the current version of {\it parADMM} is important to distribute the computations as equally as possible among the different GPU-cores and improve GPU performance
(c.f. Conclusion). The number of elements in the factor-graph grows linearly with $N$.

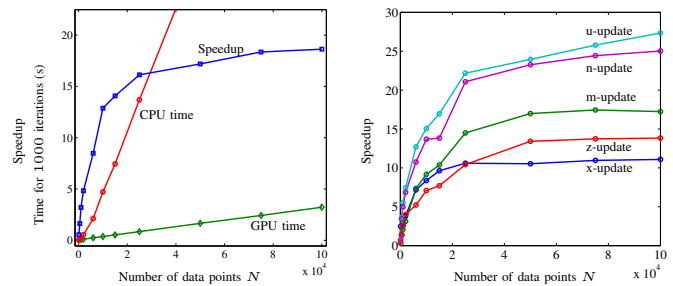
\begin{figure}[b!]
\begin{center}
\includegraphics[trim=0cm 0cm 0cm 0cm, clip=true,scale=0.27]{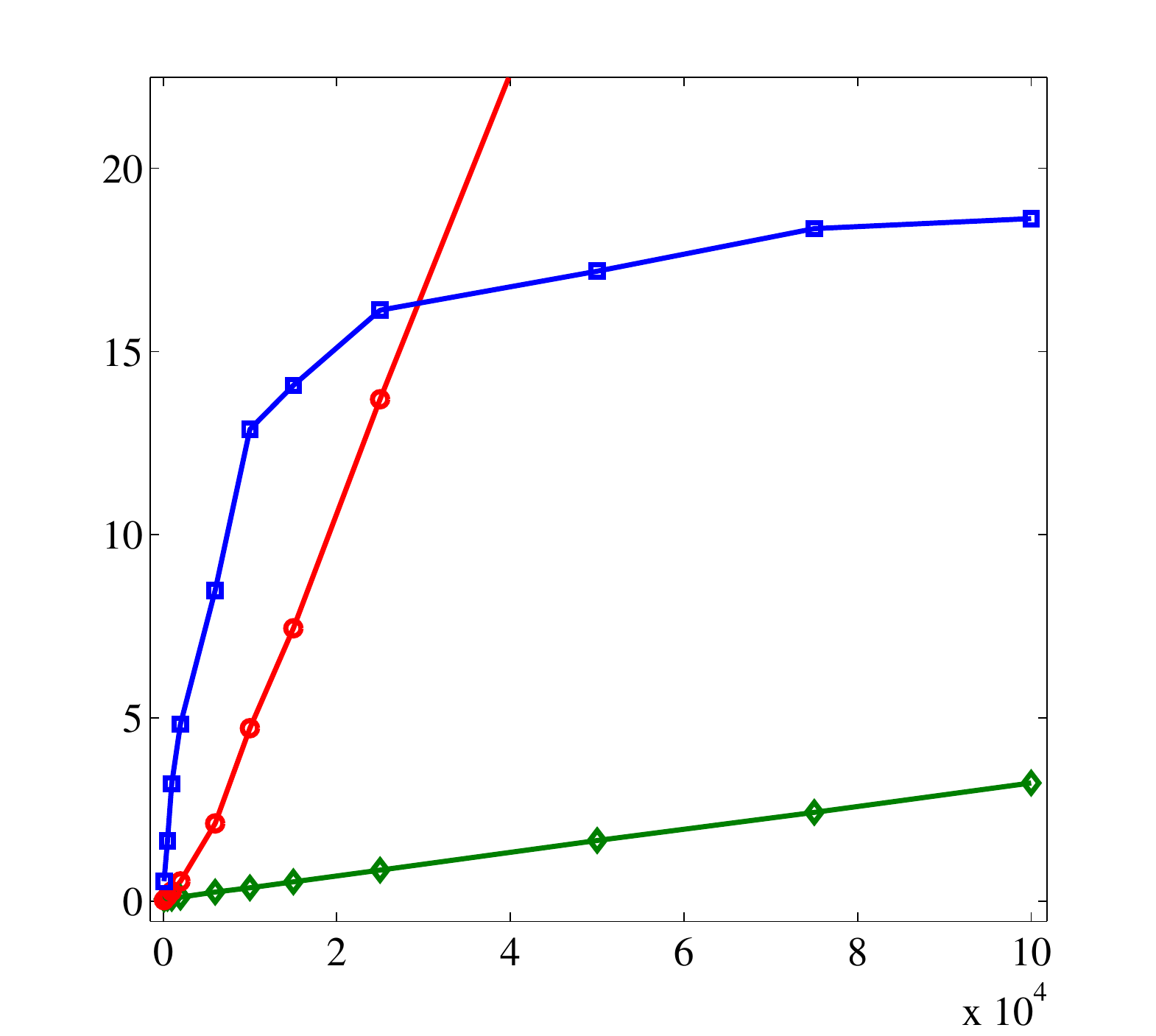}
\put(-90,-01){\tiny Number of data points $N$}
\put(-82,60){\tiny CPU time}
\put(-60,85){\tiny Speedup}
\put(-40,18){\tiny GPU time}
\put(-130,43){\tiny \rotatebox{90}{Speedup}}
\put(-122,20){\tiny \rotatebox{90}{Time for $1000$ iterations (s)}}
\includegraphics[trim=0cm 0cm 0cm 0cm, clip=true,scale=0.27]{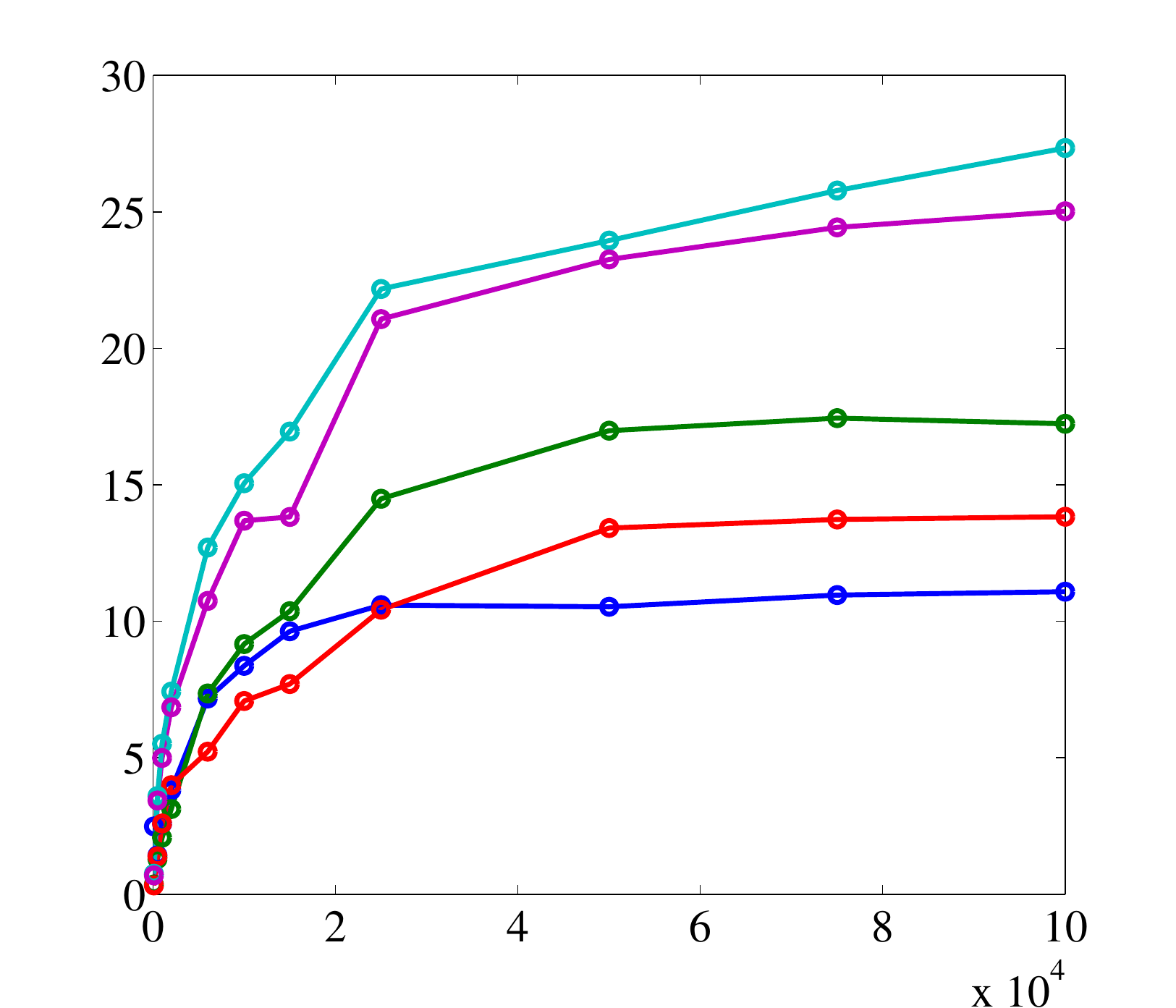}
\put(-90,-01){\tiny Number of data points $N$}
\put(-40,40){\tiny x-update}
\put(-40,48){\tiny z-update}
\put(-40,67){\tiny m-update}
\put(-40,92){\tiny u-update}
\put(-40,78){\tiny n-update}
\put(-125,43){\tiny \rotatebox{90}{Speedup}}
\caption{GPU speedup for binary classification. (Left) Combined updates; (Right) Individual updates.}
\label{fig:speed_up_SVM}
\end{center}
\end{figure}

To test the automatic speedup we can achieve we draw $N$ random data points from two Gaussian distributions with mean a certain distance apart. We do this
several times such that our results are averages over multiple random datasets. Figure \ref{fig:speed_up_SVM} summarizes our results for $x_i \in \mathbb{R}^2$.
Figure \ref{fig:speed_up_SVM}-left shows that we can achieve more than $18\times$ speedup for large problems using a GPU vs. a single CPU-core.
We also see that the time per iteration grows linearly with $N$ and hence linearly with the number of elements in the factor-graph as expected.
Figure \ref{fig:speed_up_SVM}-right shows that the individual speedups of the different kinds of updates rank in very similar order to those in circle packing and MPC, the $x$ and $z$ updates being the hardest to speedup. In our GPU experiments, the $x$ and $z$ updates again take a large fraction of the time per iteration, namely $28\% + 23\% = 51\%$.
For data $\{x_i\}$ in higher dimensions we get a lower but still substantial speedup. For example, for $N = 10^4$ and $dimension = 5,10,20,50,75,100, 150, 200$ the speedups are all between $7\times$ and $14\times$, the largest the speedup being for $dimension = 200$.

Like in packing and MPC, the time to copy the final result $z$ from the GPU to
the CPU is negligible, $60ms$ for $z \in \mathbb{R}^{2\times 10^5}$.
The time to copy the factor-graph from the CPU to the GPU is again the slowest
time, taking up to $358s$ for $N = 7.5\times 10^4$ data points.
However, just like for packing and MPC, in SVM the factor-graph is always the same for problem instances with the same number of data points. Therefore, in many practical applications, the factor-graph only needs to be copied to the GPU once and if the user wants to solve a new problem he needs only load new data onto the GPU, which must be done regardless of which method/software used.
In addition, the total time for convergence usually dominates and makes the copy time negligible.

\begin{figure}[t!]
\begin{center}
\includegraphics[trim=0cm 0cm 0cm 0cm, clip=true,scale=0.3]{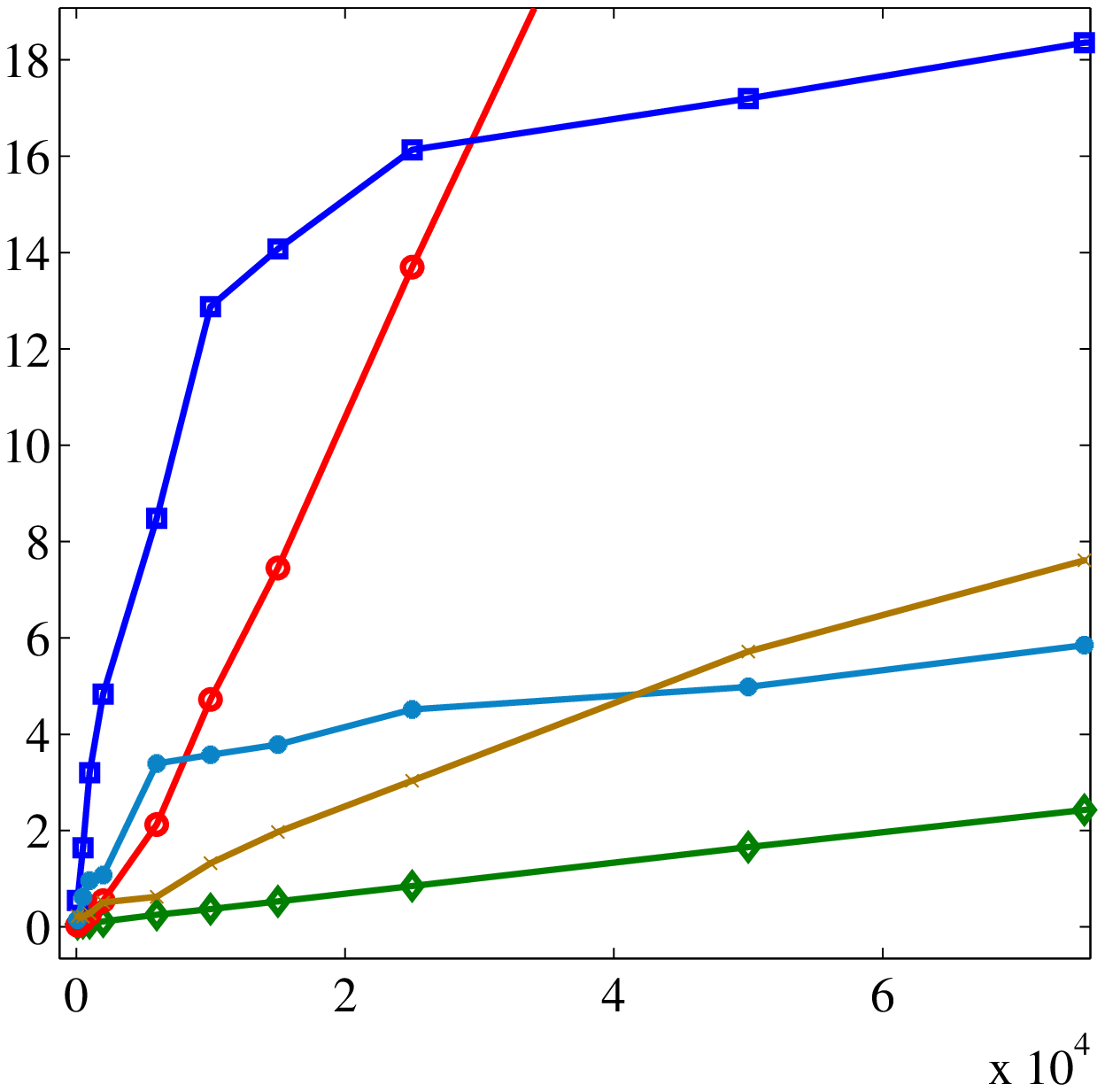}
\put(-90,-01){\tiny Number of data points $N$}
\put(-80,70){\tiny CPU time}
\put(-92,39){\tiny Multi CPU-cores speedup}
\put(-60,87){\tiny  \rotatebox{10}{GPU speedup}}
\put(-59,51){\tiny Multi CPU-cores time}
\put(-40,20){\tiny GPU time}
\put(-137,50){\tiny \rotatebox{90}{Speedup}}
\put(-127,27){\tiny \rotatebox{90}{Time for $1000$ iterations (s)}}
\includegraphics[trim=0cm 0cm 0cm 0cm, clip=true,scale=0.3]{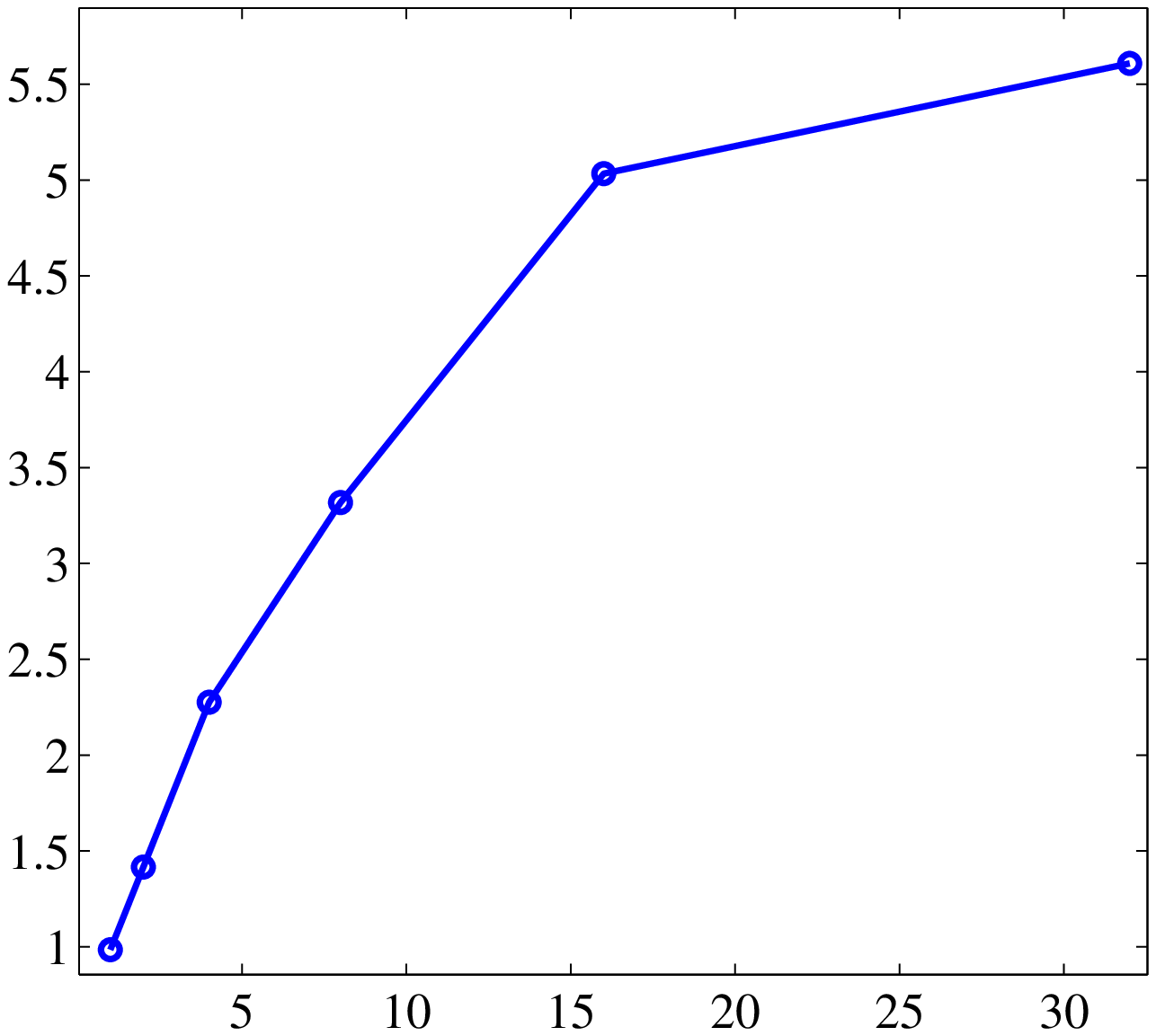}
\put(-85,-01){\tiny Number of CPU cores}
\put(-130,38){\tiny \rotatebox{90}{Multi CPU-core speedup}}
\caption{Multi CPU-cores speedup for binary classification. (Left) Combined updates for $32$ cores; (Right) Speedup vs \# threads, $N= 7.5 \times 10^4$.}
\label{fig:SVM_OpenMP}
\end{center}
\end{figure}

Figure \ref{fig:SVM_OpenMP} reports speedups using multiple CPU-cores
for $x_i \in \mathbb{R}^2$.
The right plot shows that $32$ CPU-cores provides the maximum speedup, up to $5.8\times$. In line with previous experiments, speedups using multiple CPU-cores are not as large and do not behave as well with $N$ as the GPU speedups. The individual updates are also mostly equally
heavy, taking each between $19\%$ and $25\%$ of the time per iteration.
Interestingly, for this problem, the $m$-update seems to be particularly hard to
speedup ($2.6\times$ for $N = 7.5 \times 10^4$) and the $z$-update relatively
easy to speedup ($6.2\times$ for $N = 7.5 \times 10^4$).
We note that for higher-dimensional data
we get better speedup with multiple CPU-cores (e.g. we get $9.6\times$ for $N = 10^4$, $32$ cores, $200$ dimensions).

\section{Conclusion and future work}
\noindent
We have shown that a fine-grained parallelization of the ADMM allows automatic and efficient parallelism, freeing the user from writing any
serial code while still achieving good speedups.
In the context of GPUs, our ADMM scheme is atypical in three
ways
\begin{enumerate}
\item we decompose large problems (e.g. packing) into millions of sub-problems, each assigned to a different GPU-core, instead of using the whole GPU to speedup, one-at-a-time, a few POs;
\item the GPU-cores perform tasks that are more complex and less uniform across cores than usual, each solving an optimization problem like \eqref{eq:PO};
\item our implementation only uses GPU global memory, we avoid inner-thread synchronization mechanisms by making different kernel calls for each of the different kinds of updates, and we use a very small number of threads per thread-block, making it easy to port our code to other massively parallel devices.
\end{enumerate}
We achieve speedups of up to $18\times$ using a GPU vs single CPU-core -- comparable to other GPU-accelerated libraries.

One limitation we have observed in our current implementation of this scheme is that when one GPU-core needs to perform much more work than most of the other GPU-cores, the speedup can get substantially reduced. This is in part a consequence of the fact that the $z$-update kernel only finishes once the highest-degree
variable node in the factor-graph, say $b$, is updated. Hence, since updating $z_b$ involves averaging all the $m_{(a,\partial b)}$, which is done by a single GPU-core, if the node $b$ has far more edges than most of the other variables nodes, performance can decrease. To improve this we will try a scheduling scheme
where each CUDA thread is responsible for updating not just one but several variable nodes in groups such that the total number of edges per group is as uniform as possible. Highly unbalanced degrees on the function nodes can also cause slowdowns for a similar reason.

Since the {\it parADMM} is only in its first version and is open-source, we are also counting on a collaborative effort from the broader community to test it, improve it, and extend it. We now list a few promising ideas to explore in the future.
\begin{enumerate}
\item Use asynchronous implementations of the ADMM so that not all cores need to wait for the busiest core. This is no easy task because although there are some results on asynchronous ADMM \cite{iutzeler2013asynchronous,wei20131,mota2013d} they have limitations in the topology of the graphs under which there are convergence guarantees.
\item Find algorithms that automatically adjust the factor-graph to the topology that brings greater speedups. In particular, does it always compensate to break a problem down to as many small parts as possible? Or is there is a trade-off between ending with many computationally-fast iterations and ending with a few computationally-slow iterations.
\item Extend the code to allow the use of multiple GPUs and multiple computers -- this is an easy extension but requires new code to be written.
\item In addition to improving the $z$-update step to be more robust to degree imbalance, optimize the speed of the $z$-update step in general, which on a GPU is in general the slowest one among all the updates that are not problem dependent (like the $x$-update step).
\item Test the tool on different GPUs and more CPU-cores. Some GPUs have faster clocks and more cores than the Tesla K40 use used, for example, NVIDIA's GeForce GTX TITAN X, Quadro M6000 and Tesla M40/M60. It would be interesting to understand how much hardware dependent are the speedups for different problems. In addition, in many applications floating-point precision might be enough and using cards like TITAN X might bring additional GPU speedups.
\end{enumerate}

Finally note that, although CPU clock frequencies are not expected to grow much in the future, the clock frequency of GPUs will probably still increase a bit and with certainty the number of cores per GPU will keep growing steadily. Because of this
we think that our idea to automatically exploit GPU resources in optimization using fine-grained scheduling and the ADMM will gain importance with time.

\IEEEtriggeratref{20}
\bibliographystyle{IEEEtran}
\bibliography{IEEEabrv,ref}

\newpage
\appendix
\section{Appendix for ``A general-purpose optimization tool that automatically exploits GPU parallelism''}
\noindent
In this section we include the mathematical
details regarding the computation of
all the proximal operators for the
three examples we analyze.

\subsection{Proximal operators for packing}

To solve a circle-packing problem in our framework the user has to implement serial
solvers for three different kinds of proximal operators. By exploiting symmetry we can reduce these to solving simple one-dimensional optimization problems with closed-form solutions.

The proximal operator that enforces the no-collision constraint, e.g. $\|c_1 - c_2\| \geq r_1 + r_2$, is a map that receives as input $n_{1_c},n_{1_r},n_{2_c},n_{2_r}$ and outputs the
\begin{align*}
&\arg \hspace{-4mm}\min_{c_1,r_1,c_2,r_2} \hspace{-2mm}\frac{\rho_1}{2} \| (c_1,r_1) \hspace{-1mm}-\hspace{-1mm} (n_{1_c},n_{1_r} )\|^2 \hspace{-1mm}+\hspace{-1mm}  \frac{\rho_2}{2} \| (c_2,r_2) \hspace{-1mm}-\hspace{-1mm}  (n_{2_c},n_{2_r} )\|^2\\ 
& \text{subject to } \|c_1 - c_2\| \geq r_1 + r_2.
\end{align*}
We find the solution to this problem by solving a one dimensional problem along the direction $\hat{n}=\frac{n_{2_c} - n_{1_c}}{\|n_{2_c} - n_{1_c}\|}$. The final solution is them
\begin{align*}
&(c_1,r_1) = (n_{1_c},n_{1_r}) + \frac{D}{2}\frac{\rho_2}{\rho_1 + \rho_2} (-\hat{n},1)\\
&(c_2,r_2) = (n_{2_c},n_{2_r}) + \frac{D}{2}\frac{\rho_1}{\rho_1 + \rho_2} (\hat{n},1)\\
&\text{where }D = \max\{0,n_{r_1}+n_{r_2} - \|n_{c_1} -n_{c_2} \| \}.
\end{align*}

The proximal operator that enforces no collisions with a wall, specified by a plane with normal direction $Q$ and a point $V$, is a map that receives as input $n_{c},n_{r}$ and outputs the
\begin{align*}
&\arg \min_{c, r} \|(c,r) - (n_c,n_r)\|^2\\
&\text{subject to } Q^\dagger(c - V) \geq r.
\end{align*}
The solution to this problem is
\begin{align*}
&(c,r) = (n_c,n_r) + E(-Q,1)\\
&\text{where } E = \min\{0,\frac{1}{2}\left(Q^\dagger(n_c - V) - n_r\right)\}.
\end{align*}

Finally, the PO that tries to maximize the radius of
each sphere is a map that receives as input $n_{r}$ and outputs the
\begin{align*}
\arg \min_{r} -\frac{1}{2} r^2 + \frac{\rho}{2}(r - n_r)^2 = \frac{\rho}{-1 + \rho} n_r.
\end{align*}

\subsection{Proximal operators for MPC}
We have used two different proximal operators for the MPC problem; (1) Cost Function (2) Linearized System Dynamic.

The cost function proximal operator is defined by
\begin{align*}
& \arg \min_{x(i)} \sum_{i=0}^{K-1} x^T(i) Q x(i) + \sum_{i=0}^{K} u(i)^T R u(i) + x^T(K)Q_fx(K) \\
& + \sum_{i=0}^{K} \frac{\rho (i)}{2} \big( \left\| x(i) - n_{x}(i) \right\| ^2 + ( u(i) - n_{u}(i) )^2 \big)
\end{align*}
for all $i=0,\ldots,K$. Note that there is a closed form solution for this which can be obtained as
\begin{align*}
&\begin{bmatrix}x(i) \\ u(i) \end{bmatrix} = \rho(i) \begin{bmatrix} (Q+\rho(i)I)^{-1} && 0 \\ 0 && (R + \rho(i)I)^{-1} \end{bmatrix} \begin{bmatrix} n_{x}(i) \\ n_{u}(i) \end{bmatrix}
\end{align*}
\noindent The linearized system dynamic is
\begin{align*}
& \arg \min_{x(i),x(i+1)} \frac{\rho(i)}{2} \big( \left\| x(i) - n_x(i) \right\|^2 + (u(i) - n_u(i) )^2 \big) \\
& + \frac{\rho(i+1)}{2} \big( \left\| x(i+1) - n_x(i+1) \right\|^2 + (u(i+1) - n_u(i+1) )^2 \big) 
\end{align*}
for all $i=0,\ldots,K$. This can also be solved in closed form.

\subsection{Proximal operators for SVM}
As explained, we use three types of proximal operators to solve SVM training optimization, and then couple them using the ADMM factor graph.

\subsubsection{Minimal Error Proximal  Operator}
This proximal operator reads as:

\begin{equation} \label{5}
\begin{split}
 &\hat{\xi}=\arg\min_{\xi} \sum_{i=1}^n\frac{\rho_i}{2}(\xi_i-n_i)^2 +\lambda \sum_{i=1}^n\xi_i\\
&0\le\xi_i.
\end{split}
\end{equation}
Basically, it is a semi-lasso problem and it has closed form solution as:
\begin{equation} \label{6}
\begin{split}
\hat{\xi_i}=\left(n_i-\frac{\lambda}{\rho_i}\right)^+.
\end{split}
\end{equation} 
 
\subsubsection{Minimal Norm Two Proximal  Operator}
This proximal operator reads as:
\begin{equation} \label{7}
\begin{split}
 &\hat{w}=\arg\min_{w} \frac{1}{2}
 \|w\|^2+ \frac{\rho}{2}\|w-n\|^2.
\end{split}
\end{equation}
It is a least square problem can easily be solved in closed form:
\begin{equation} \label{8}
\begin{split}
\hat{w}=\frac{\rho}{\rho+1}n
\end{split}
\end{equation}
 \subsubsection{One Point Minimal Margin Proximal Operator}
 This proximal operator reads as:
 \begin{equation} \label{9}
 \begin{split}
  &\{\hat{w},\hat{\xi},\hat{b}\}=\arg\min_{w,\xi,b} \frac{\rho_1}{2}\|w-n_1\|^2+\frac{\rho_2}{2}(b-n_2)^2\\&+\frac{\rho_3}{2}( \xi - n_3)^2\\
 &\text{s.t.}  y (w.x +b)\ge 1-\xi_,
 \end{split}
 \end{equation}
 which is a least square objective subject to a half-plane constraint. Using Lagrange multipliers, it can be solved in closed form as:
 \begin{equation} \label{10}
  \begin{split}
& \alpha=\left( \frac{y(n_1.w+n_2b)+n_3-1}{\frac{\|x\|^2}{\rho_1}+\frac{1}{\rho_2}+\frac{1}{\rho_3}}\right)^+\\
&\hat{w}=n_1-\frac{\alpha}{\rho_1}yx
\\
&\hat{b}=n_2-\frac{\alpha}{\rho_2}
\\
&\hat{\xi}=n_3-\frac{\alpha}{\rho_3}.
  \end{split}
  \end{equation}
 \subsubsection{Equality Proximal Operator}
 This auxiliary proximal operator reads as:
   \begin{equation} \label{11}
   \begin{split}
    &\{\hat{w}_1,\hat{w}_2\}=\arg\min_{w_1,w_2} \frac{\rho_1}{2}\|w_1-n_1\|^2+\frac{\rho_2}{2}\|w_2-n_2\|^2\\
   &\text{s.t.} w_1=w_2 ,
   \end{split}
   \end{equation}
   which  can be solved in closed form as:
   \begin{equation} \label{12}
    \begin{split}
  w_1=w_2=\frac{\rho_1n_1+\rho_2 n_2}{\rho_1+\rho_2}
   \end{split}
    \end{equation}
Given these proximal operators, we can couple them using the factor graph represented in Figure \ref{fig:SVM_formulation_and_graph} and then by message-passing iteratively solve SVM optimization problem via our proposed toolbox.

\end{document}